\documentclass[12pt]{article}

\usepackage{graphicx}
\usepackage{natbib}
\usepackage[usenames,dvipsnames]{color}
\usepackage{amssymb}
\usepackage{amsmath} 
\usepackage{mathrsfs}

\usepackage{setspace}

\usepackage[font = {it}]{caption}

\newtheorem{lemma}{Lemma}

\newtheorem{proposition}{Proposition}
\newtheorem{corollary}{Corollary}

\RequirePackage{bm}

\newcommand{\dd}{{\rm d}}

\usepackage{amssymb}
\usepackage{graphicx}
\usepackage{ulem}

\title{Decompositions of Dependence for High-Dimensional Extremes}

\author{Daniel Cooley\\
Colorado State University
\vspace{.2 in}\\
Emeric Thibaud\\
Ecole Polytechnique F\'ed\'erale de Lausanne}

\begin{document}
\maketitle


\begin{abstract}
Employing the framework of regular variation, we propose two decompositions which help to summarize and describe high-dimensional tail dependence. 
Via transformation, we define a vector space on the positive orthant, yielding the notion of basis. 
With a suitably-chosen transformation, we show that transformed-linear operations applied to regularly varying random vectors preserve regular variation. 
Rather than model regular variation's angular measure, we summarize tail dependence via a matrix of pairwise tail dependence metrics. 
This matrix is positive semidefinite, and eigendecomposition allows one to interpret tail dependence via the resulting eigenbasis.
Additionally this matrix is completely positive, and a resulting decomposition allows one to easily construct regularly varying random vectors which share the same pairwise tail dependencies.  
We illustrate our methods with Swiss rainfall data and financial return data.
\end{abstract}


%
%
%
%
%
%
%
%
%
%
%
%
%


\allowdisplaybreaks

\section{Introduction}

Despite many noteworthy models which have been developed to capture extremal dependence like the extremal-$t$ \citep{opitz2013}, H\"usler--Reiss (\citeyear{husler1989}), and mixture-Dirichlet \citep{boldi2007},
describing dependence for high-dimensional extremes remains a challenge.
Multivariate regular variation is intricately linked to the study of extremes as it can be tied to characterizations of the multivariate extreme value distributions or used directly to model threshold exceedances.
Tail dependence of a multivariate regularly varying random vector of dimension $p$ is described by the angular (or spectral) measure on the $p-1$ dimensional unit ball.
Estimating a $p-1$ dimensional measure is challenging, and this challenge is exacerbated as extremes practice retains only a small subset of extreme data for estimation.
Recent efforts to describe or model high dimensional extremes include 
\cite{strokorb2013} who determine a max-stable model from pairwise and higher-dimensional summaries of extremal dependence, and \cite{chautru2015} who proposes a clustering approach to group variables.

In the non-extreme setting, the covariance matrix or its estimate is widely used to summarize dependence information for multivariate distributions or data, even when non-Gaussian.
Many multivariate methods arise by performing linear operations on the sample covariance matrix.
For example, principal component analysis \citep[][Ch. 8]{johnson2007} is performed by an eigendecomposition of the covariance matrix. 
We aim to provide tools for exploring extremal dependence in high dimensions. 
We will work within the framework of multivariate regular variation, 
producing a matrix which summarizes tail dependence, and ultimately obtain two useful decompositions of this matrix.
The first decomposition provides an ordered orthonormal basis useful for examining the modes of extremal dependence.
The second provides a method for constructing a random vector with a simple dependence structure, yet which shares the same pairwise tail dependencies.
To produce these decompositions we link the seemingly disparate ideas of inner product spaces to multivariate regular variation on the nonnegative orthant, thus enabling us to perform transformed linear operations on multivariate regularly varying random vectors. 

\cite{resnick2007} gives a comprehensive treatment of multivariate regular variation. 
Roughly speaking, a random vector is multivariate regularly varying if its joint tail decays like a power function; i.e., it is jointly heavy-tailed.
Although regular variation can be defined on $\mathbb{R}^p$ \citep[\S6.5.5]{resnick2007}, we model in the nonnegative orthant, as this allows one to focus attention to the direction for which one wants to assess risk.
Let $X$ be a random vector which takes values in $\mathbb{R}^p_+ = [0, \infty)^p$.
The vector $X$ is regularly varying if there exists a sequence $ b_n \rightarrow \infty$ and a limit measure $\nu_{X}$ for sets in $[0, \infty]^p \setminus \{0\}$ such that $n \Pr( b_n^{-1}{X} \in \cdot) \stackrel{v}{\rightarrow} \nu_{X} (\cdot)$, as $n \rightarrow \infty$, where $\stackrel{v}{\rightarrow}$ denotes vague convergence in $M_+([0, \infty]^p \setminus \{ 0 \})$, the space of nonnegative Radon measures on $[0, \infty]^p \setminus \{  0 \}$ \citep[\S6]{resnick2007}. 
It can be shown that ${b_n} = L(n) n^{1/\alpha}$, where $L(n)$ is some slowly varying function and $\alpha >0$ is called the tail index of $X$. We denote $X \in RV_+^p(\alpha)$ a regularly varying vector $X$ with tail index $\alpha$. 
The measure $\nu_{X}$ has the scaling property $\nu_{X}(a C) = a^{-\alpha} \nu_{X}(C)$ for any set $C \subset [0, \infty]^p \setminus \{ 0 \}$ and any $a > 0$.

The scaling property implies that $\nu_{X}$ can be more easily understood for sets defined by polar, rather than Cartesian, coordinates.
Given any norm, define the unit ball $\mathbb{S}^+_{p-1} =  \{ x \in \mathbb{R}^p_+ : \| x\| = 1\}$.
Let $C(r, B) = \{ x \in \mathbb{R}^p_+ : \| x \| > r,  \| x \|^{-1} x \in B\}$ for some $r >0$, and some Borel set $B \subset \mathbb{S}^+_{p-1}$.
Then $\nu_{X}\{C(r,B)\} = r^{-\alpha} H_{X}(B)$, where $H_{X}$ is a measure, termed the angular measure, on $\mathbb{S}^+_{p-1}$.  
Consequently, $\nu_{X}(\dd r \times \dd w) = \alpha r^{-\alpha - 1} \dd r \dd H_{X}(w)$.  
The scale of $X$ is related to $\{b_n\}$ and $\nu_X$ or $H_X$.
Consider replacing the sequence $\{b_n\}$ by $\{k b_n\}$ for some $k > 0$. 
For any $r>0$ and $B \subset \mathbb{S}^+_{p-1}$ such that $C(r,B)$ is a continuity set of $\nu_{X}$,
$
  n \Pr\{ (k b_n)^{-1}X \in C(r, B)\}
  \rightarrow 
\nu_{X}^{(alt)} \{C(r, B)\} = r^{-\alpha} H_{X}^{(alt)}(B),
$
where $\nu_{X}^{(alt)} = k^{-\alpha} \nu_{X}$ and $H_{X}^{(alt)} = k^{-\alpha} H_{X}$.
We will acknowledge this scaling relationship between $\{b_n\}$, $\nu_{X}$, and $H_{X}$ by saying that $X$ has limiting (angular) measure $\nu_{X}$ (respectively, $H_{X}$) when normalized by $\{b_n\}$.


\section{Inner product space via transformation}
\label{sec: innerProduct}
In this section, we describe a framework to produce an inner product space on a target open set, yielding, among other things, the notion of basis.
We later use this framework to produce a particular inner product space on the positive orthant whose operations will preserve regular variation.
Let $t$ be a bijection from $\mathbb{R}$ onto some open set $\mathbb{X}$, and let $t^{-1}$ be its inverse.
We refer to $t$ as the `transform'.
Define $\mathbb{X}^p$ as the set of $p$-dimensional vectors whose elements are in $\mathbb{X}$.
Let $t(y)$ denote element-wise application of $t$ to the elements of $y\in\mathbb{R}^p$, and other functions applied to vectors will similarly be applied element-wise.  
Define vector addition in $\mathbb{X}^p$ as $x_1 \oplus x_2 = t\big\{  t^{-1}(x_1) + t^{-1}(x_2) \big\}$.
For $a \in \mathbb{R}$, define scalar multiplication of a vector in $\mathbb{X}^p$ as $a \circ x = t\big\{ a t^{-1}(x) \big\}$.
Define the additive identity in $\mathbb{X}^p$, $0_{\mathbb{X}^p} = t(0)$, and define the additive inverse of any $x \in \mathbb{X}^p$ as $- x = t\big\{ - t^{-1} (x)\big\}$.
We show that $\mathbb{X}^p$ meets the ten conditions required of a vector space in the supplemental materials.

Continuing to let $x_j \in \mathbb{X}^p$ and $a_j \in \mathbb{R}$ $(j = 1, \ldots, q)$, we define a linear combination in the obvious way:
\begin{equation*}
a_1 \circ x_1 \oplus \cdots \oplus a_q \circ x_q = t \Bigg\{ \sum_{j = 1}^q a_j t^{-1} (x_{j}) \Bigg\}.
\end{equation*}
As $\mathbb{X}^p$ is a $p$-dimensional vector space, any set of $p$ vectors which are linearly independent in $\mathbb{X}^p$ (i.e., if $a_1 \circ x_1 \oplus \cdots \oplus a_p \circ x_p = 0 \Rightarrow a_1 = \cdots = a_p = 0$) form a basis for $\mathbb{X}^p$.

Let $A = (a_{1}, \ldots, a_{q})$ be a $p \times q$ matrix of real numbers. 
For $x \in \mathbb{X}^q$, define matrix multiplication as $A \circ x = a_{1} \circ x_{1} \oplus \cdots \oplus a_{q} \circ x_{q}
= t\{A t^{-1} (x)\}$, and note $A \circ x \in \mathbb{X}^p$. 
If $I$ is the usual $p$-dimensional identity matrix, then $I \circ x = t\big\{I t^{-1} (x) \big\} = x$.
Also note that if $B$ is $p' \times p$, then $B \circ A \circ x = B \circ t\{At^{-1}(x)\} = t\{BA t^{-1}(x)\} = BA \circ x$.

As in $\mathbb{R}^p$, linear combinations can be written as matrix operations.
However, because the constants $a_i \in \mathbb{R}$ and the vectors $x_i \in \mathbb{X}^p$, the relationship changes slightly:
\begin{equation}
  a_1 \circ x_1 \oplus \cdots \oplus a_q \circ x_q 
%
  = Y \circ t(a) \label{eq: linCombo},
\end{equation} 
where $Y\in \mathbb{R}^{p \times q}$ is the matrix whose columns are $y_i = t^{-1}(x_i)$ $(i = 1, \ldots, q)$.

For $x_1, x_2 \in \mathbb{X}^p$, define the scalar product $\langle x_1, x_2 \rangle = \sum_{i = 1}^p t^{-1} (x_{1i}) t^{-1} (x_{2i})$.
In the supplemental materials, we show that the four conditions of a scalar product are met.
Define $\| x \| = \langle x, x \rangle^{1/2}$, and 
define two vectors to be orthogonal if $\langle x_1, x_2 \rangle = 0$, denoted $x_1 \bot x_2$.  
Vectors $x_1, x_2 \in \mathbb{X}^p$ and their preimages $y_1 = t^{-1} (x_1), y_2 = t^{-1} (x_2) \in \mathbb{R}^p$ share the same inner product value.
Consequently, $\| x \| = \| y \|_2$, and 
$x_1 \bot x_2$ in $\mathbb{X}^p$ if and only if $y_1 \bot y_2$ in $\mathbb{R}^p$.


Consider nonsingular matrix $S \in \mathbb{R}^{p \times p}$, and think of $S$ as an operator $\mathbb{X}^p \rightarrow \mathbb{X}^p$ defined by $x\mapsto S \circ x$.
We define the inverse operator $S^{-1}$ to be a matrix such that $S^{-1} \circ \big( S\circ x\big) = S \circ \big (S^{-1} \circ x \big) = x$, and note that the inverse operator coincides with the usual matrix inverse.
Define an eigenvalue/eigenvector pair $\lambda \in \mathbb{R}$, $ e \in \mathbb{X}^p$ of $S$ to be such that $S \circ  e = \lambda \circ  e$, and assume $\| e \| = 1$.
If $\lambda$ and $ u$ are an eigenvalue/eigenvector pair in $\mathbb{R}^p$, then $\lambda$ and $ e = t( u)$ is an eigenvalue/eigenvector pair in $\mathbb{X}^p$:
\begin{equation*}
S \circ  e  = t(S u) = t( \lambda  u) = \lambda \circ  e.
\end{equation*}

Further, assume that $S$ is symmetric positive-definite (i.e., $ y^T S  y > 0$ for any $ y \in \mathbb{R}^p\setminus\{0\}$), and
define a positive quadratic form  in $\mathbb{X}^p$ as $Q(S,  x) = \langle  x, S \circ  x \rangle$. 
Note that
\begin{equation*}
  Q(S,  x) 
  = \big \langle  x, t\big\{ S t^{-1}( x) \big\} \big \rangle 
  = \sum_{i = 1}^p \sum_{j = 1}^p t^{-1}(x_i) s_{i j} t^{-1} (x_j)  
=   y^T S  y,
\end{equation*}
where $ y = t^{-1} ( x)$.
Thus, $ x$ and its preimage $ y$ share the same quadratic form with symmetric positive definite matrix $S$. 
Consequently, relationships between the eigenvectors and eigenvalues of $S$ and bounds on the quadratic forms in $\mathbb{R}^p$ carry over to $\mathbb{X}^p$ yielding the following proposition, whose justification follows from linear algebra results in $\mathbb{R}^p$ \citep[p. 80]{johnson2007}.

\begin{proposition}
  \label{prop: eigOrder}
Let $\lambda_1 \geq \cdots \geq \lambda_p > 0$ and $ u_1, \ldots,  u_p$ be the ordered eigenvalue/eigenvector pairs for $S$ in $\mathbb{R}^p$, and let $ e_i = t( u_i)$ $(i = 1, \ldots, p)$.
Then, 
\begin{align*}
  \max_{ x: \|  x\| = 1} Q(S,  x) &= \lambda_1, \mbox{ occurring when }  x =  e_1 \mbox{, and}\\
  \min_{ x: \|  x\| = 1} Q(S,  x) &= \lambda_p, \mbox{ occurring when }  x =  e_p.
\end{align*}
Further, the sequence of vectors $ x_1,  x_2, \ldots,  x_p$ where each $ x_k$ is such that $Q(S,  x_k)$ is maximized subject to $ x_k \bot   x_i$, $i < k$, is $ x_i =  e_i$ $(i = 1, \ldots, p)$ and $Q(S,  e_i) = \lambda_i$.
\end{proposition}

In~\S\ref{sec: linOpsRegVar}, we apply the ideas of this section with the specific transform $t: \mathbb{R} \rightarrow (0, \infty)$ defined as 
\begin{equation*}
  \label{eqn:transform}
  t(y) = \log\{1 + \exp(y)\}.
\end{equation*}
This bijection, known as the softplus function in neural networks \citep{dugas2001}, is continuous and infinitely differentiable. 
Its inverse is $t^{-1}(x) = \log\{\exp(x) - 1\}$. 
Importantly for our purposes,
$\lim_{y \rightarrow \infty} t(y)/y = \lim_{x \rightarrow \infty} t^{-1}(x)/x = 1$; 
that is, the transform and its inverse negligibly affect large values.  
For use in~\S\ref{sec: linOpsRegVar}, extend $t$ such that $t(-\infty) = 0$, $t^{-1}(0) = -\infty$, and $t(\infty) = t^{-1}(\infty) = \infty$.
As defined, $t: \bar{\mathbb{R}}^p \rightarrow \bar{\mathbb{X}}^p$, where $\bar{\mathbb{R}}^p = [-\infty, \infty]^p$ and $\bar{\mathbb{X}}^p = [0, \infty]^p$. 
The additive zero vector in $\bar{\mathbb{X}}^p$ is a vector of $p$ components $t(0) = \log 2$.

\section{Transformed-linear operations on multivariate regularly varying random vectors}
\label{sec: linOpsRegVar}





We now consider  `transformed-linear' operations applied to regularly varying random vectors with the aforementioned specific transform $ t(y) = \log\{1 + \exp(y)\}$. 
For regular variation to be preserved we need the following assumption on the lower tail of our random vectors: 
assume $ X \in RV_+^p(\alpha)$ verifies 
\begin{equation}\label{eq:lowertailcond}
n \Pr\{X_i \leq \exp(-k b_n)\} \rightarrow 0,  \quad k>0, i=1,\ldots,p,
\end{equation}
as $n \rightarrow \infty$.
Condition~\eqref{eq:lowertailcond} does not seem overly restrictive as $\exp(-k b_n)$ goes to zero very rapidly, but it does preclude any of the marginals from having nonzero mass at 0.  
Standard regularly varying distributions such as the Pareto and the Fr\'echet meet this condition.
%

Propositions \ref{prop: sum} and \ref{prop: mult} below show that regular variation is preserved by the transformed linear operations.  
Proofs for all propositions are given in the appendix.  
Used to prove Propositions  \ref{prop: sum} and \ref{prop: mult}, Lemmas A\ref{lem:tm1XisRV} and A\ref{lem:tYisRV} require the definition of regular variation on $\mathbb{R}^p$, which we recall here and which is used in Section \ref{sec: eigenDecomp}.
$Y$ is regularly varying in $\mathbb{R}^p$ (denoted as $ Y \in RV^p(\alpha)$)  if there exists $b_n \rightarrow \infty$ such that $n \Pr( b_n^{-1}  Y \in \cdot)  \stackrel{v}{\rightarrow} \nu_{ Y}(\cdot)$ in $M_+(\bar{\mathbb{R}}^p\setminus\{ 0\})$ \citep[][\S6.5.5]{basrak2002,resnick2007}.
Analogous to before, if $C(r,B) = \{ y \in \mathbb{R}^p : \|y\| > r, \|y\|^{-1}y \in B\}$ for some set $B \subset \mathbb{S}_{p-1} = \{y \in \mathbb{R}^p : \|y \| = 1\}$, then $\nu_Y\{C(r,B)\} = r^{-\alpha} H_Y(B)$ for some angular measure $H_Y$ on $\mathbb{S}_{p-1}$.
Notationally for $ y\in\mathbb{R}^p$,  let $ y^{(0)} = \max( y,  0)$ applied componentwise.

\begin{proposition} 
  \label{prop: sum}
Let $ X_1,  X_2 \in RV_+^p(\alpha)$ be independent, 
$n \Pr(b_n^{-1}  X_1 \in \cdot) \stackrel{v}{\rightarrow} \nu_{ X_1}(\cdot)$, 
and 
$n \Pr(b_n^{-1}  X_2 \in \cdot) \stackrel{v}{\rightarrow} \nu_{ X_2}(\cdot)$. 
Then $ X_1 \oplus  X_2 = t \big\{ t^{-1} ( X_1) + t^{-1} ( X_2) \big\} \in RV_+^p(\alpha)$, and
\begin{equation*}
n\Pr\{ b_n^{-1} (X_1 \oplus  X_2) \in \cdot\} \stackrel{v}{\rightarrow} \nu_{ X_1}(\cdot) + \nu_{ X_2}(\cdot).
\end{equation*}
\end{proposition}

\begin{proposition} 
  \label{prop: mult}
  Let $ X \in RV_+^p(\alpha)$ be such that $n\Pr(b_n^{-1}  X \in \cdot) \stackrel{v}{\rightarrow} \nu_{ X}(\cdot)$.  
  Then for $a \in \mathbb{R}$,
  \begin{align*}
    n \Pr \{
    		b_n^{-1}(a \circ  X) \in \cdot 
	  \} 
    &\stackrel{v}{\rightarrow} a^\alpha \nu_{ X}( \cdot ) \mbox { if } a > 0, \mbox{ and }\\
    n \Pr \{
    		b_n^{-1}(a \circ  X) \in \cdot 
	  \} 
    &\stackrel{v}{\rightarrow} 0 \mbox { if } a \leq 0.
  \end{align*}
\end{proposition}
Informally, the condition~\eqref{eq:lowertailcond} is necessary because if $ X$ had enough mass near $ 0$ and $a < 0$,  $t\{at^{-1}( X)\}$ could interfere with the regular variation in the upper tail.  

One outcome of propositions \ref{prop: sum} and \ref{prop: mult} is a method for constructing a regular varying random vector by applying a matrix $A$ to a vector of independent regularly varying random variables.  

\begin{corollary}
  \label{coro: mtxMult}
  Let $A = ( a_{1}, \ldots,  a_{q})$ be a $p \times q$ matrix
  where $\max_{i = 1, \ldots, p} a_{ij} > 0$ for all $j = 1, \ldots q$, 
  and let $ Z = (Z_1, \ldots, Z_q)^T$ be a vector of independent and identically distributed regularly varying $\alpha$ random variables with $\{b_n\}$ such that $n \Pr( Z_j > b_n z ) \rightarrow z^{-\alpha}$, $j = 1, \ldots, q$, and $n \Pr\{Z_j \leq \exp(-k b_n)\} \rightarrow 0$ for any $k > 0$, $j = 1, \ldots, q$.  
  Then $A \circ  Z  \in RV_+^p(\alpha)$ and when normalized by $\{b_n\}$ has angular measure  
    $H_{A \circ  Z}(\cdot) = \sum_{j = 1}^q \|  a^{(0)}_{j} \|^\alpha 
	\delta_{ a^{(0)}_{j} / \|  a^{(0)}_{j} \|}(\cdot)$,
  where $\delta$ is the Dirac mass function. 
\end{corollary}
Different matrices can result in the same limiting angular measure. 
Let $A^{(0)} = (a_{ij}^{(0)})_{i = 1, \ldots p; j = 1, \ldots, q}.$
If $A \neq A'$ but $A^{(0)} = A'^{(0)}$, $H_{A \circ  Z} = H_{A' \circ  Z}$.

The class of regularly varying random vectors produced by the construction method in Corollary \ref{coro: mtxMult} is similar to the family of random vectors defined by max-linear combinations of independent regularly varying random variables \citep[e.g.,][]{schlather2002, fougeres2013}.
Let $A$ be a $p \times q$ matrix and let $ Z = (Z_1, \ldots, Z_q)^T$ be a vector of independent and identically distributed regularly varying $\alpha$ random variables  as in Corollary~\ref{coro: mtxMult}.
Constructing 
$
  A \times_{\max}  Z = \left(\max_{j = 1,\ldots, q} a_{1j} Z_j, \ldots, \max_{j = 1,\ldots, q} a_{pj} Z_j\right)^T,
$
one can show that $H_{A \times_{\max}  Z} = H_{A \circ  Z}$.
If $ Z$ is max-stable, then $A \times_{\max}  Z$ is also max-stable.
We choose to work within the framework of regular variation rather than max-stability, but given knowledge of the angular measure it is easy to determine the distribution of the max-stable random vector whose domain of attraction includes the regularly varying random vector.
One difference between the two constructions is in their realizations.
Large realizations of the max-linear combination tend to have angular components which correspond exactly to the discrete locations for which the angular measure has mass, whereas large realizations of the transformed-linear construction have angular components close but not equal to these discrete locations.
  


Similar to \cite{fougeres2013} who show that the class of max-linear combinations of independent Fr\'echet random variables is dense in the class of $p$-dimensional multivariate extreme value distributions with Fr\'echet marginals, Proposition \ref{prop: dense} below shows the class of angular measures arising from the construction method in Corollary \ref{coro: mtxMult} is dense in the class of possible angular measures.
To construct the dense class one only needs to consider matrices $A$ with nonnegative elements, and the approximation to a continuous angular measure is improved by increasing the number of columns of $A$.

\begin{proposition}
\label{prop: dense}
  Given any angular measure $H$, there exists a sequence of nonnegative matrices $\{A_q\}$, $q = 1, 2, \ldots$, such that $H_{A_q \circ  Z_q} \stackrel{w}{\rightarrow} H$.
\end{proposition}  

\section{Tail pairwise dependence matrix}
\label{sec: TPDM}

Inspired by statistical practice in non-extreme settings, we aim to summarize dependence of a regularly varying random vector via second-order properties of its angular measure.
Henceforth, we restrict our attention to the case $\alpha = 2$, and to employ the $L_2$ norm when making the radial/angular transformation.

Assume $ X \in RV_+^p(2)$ such that 
\begin{equation}
  \label{eq: restrictedX}
  n \Pr( n^{-1/2} X \in \cdot ) \stackrel{v}{\rightarrow} \nu_{ X} (\cdot), \mbox{ where } \nu_{ X}(\dd r \times \dd w) = 2 r^{-3} \dd r \dd H_{ X} ( w),
\end{equation}
and $H_{ X}$ is a Radon measure on the $L_2$ unit ball $\Theta^+_{p-1} = \{  w \in \bar{\mathbb{X}}^p : \|  w \|_2 = 1 \}$. 
We have specified the normalizing sequence to be $\{n^{1/2}\}$, thus pushing all scaling information into~$H_{ X}$.
If one begins with a random vector in $ RV_+^p(\alpha)$, then a marginal transformation can be applied to achieve a random vector which meets the above conditions \citep[][Theorem 6.5]{resnick2007}.

Define a $p \times p$ matrix of summary pairwise dependencies by letting
\begin{equation*}
  {\sigma_{ X}}_{i k} = \int_{\Theta^+_{p-1}} w_i w_k \dd H_{ X} ( w), \mbox{ and } \Sigma_{ X} = ({\sigma_{ X}}_{i k})_{i ,k = 1, \ldots, p}.
\end{equation*}
We refer to $\Sigma_{ X}$ as the tail pairwise dependence matrix of $X$, and ${\sigma_{ X}}_{i k}$ corresponds to the extremal dependence measure, defined in the bivariate case by \cite{larsson2012}.

It is straightforward to show that $\Sigma_{ X}$ is positive semidefinite. 
Let $m = H_X( \Theta_{p-1}^+ )$. 
Let $ W$ be a random vector such that $\Pr( W \in B) = m^{-1} H_{ X}(B)$ for any set $B \subset \Theta^+_{p-1}$. 
Then $\Sigma_{ X} = m {\rm E}(  W  W^T )$, and $ y^T \Sigma_{ X}  y = m {\rm E}(  y^T  W  W^T  y ) \geq 0$, for $y \in \mathbb{R}^p\setminus\{0\}$.
The inequality becomes strict if no element of $ W$ is a linear combination of the others.

Like a covariance matrix, the diagonal elements of $\Sigma_{ X}$ yield information about the scale of the elements of $ X$.
Since $\alpha = 2$, 
\begin{equation*}
  \lim_{n \rightarrow \infty} n \Pr( n^{-1/2} X_i > x) = \int_{\Theta^+_{p-1}} \int_{x/w_i}^\infty 2r^{-3} \dd r \dd H_{ X}( w) = x^{-2} \int_{\Theta^+_{p-1}} w_i^2 \dd H_{ X}( w) = x^{-2} {\sigma_{ X}}_{ii}.
\end{equation*}
Thus, ${\sigma_{ X}}_{ii}$ is equal to the square of the scale of $X_i$, since if $X_i$ has scale 1 (defined as: $\lim_{n \rightarrow \infty} n \Pr( n^{-1/2}{X_i} > x ) = x^{-2}$), then $\lim_{n \rightarrow \infty} n \Pr( n^{-1/2} {cX_i} > x) = c^2 x^{-2}$.
Also since we use the $L_2$ norm, the sum of the diagonal elements is equal to the total mass of the angular measure as
\begin{equation*}
  \label{eqn: totalMass}
  \sum_{i = 1}^p {\sigma_{ X}}_{ii} = \int_{\Theta^+_{p-1}} \sum_{i = 1}^p w_i^2 \dd H_{ X}( w) = \int_{\Theta^+_{p-1}} \dd H_{ X}( w).
\end{equation*}

Asymptotic independence \citep{sibuya1960, ledford1996} of the components $X_i$ and $X_k$ is equivalent to ${\sigma_{ X}}_{ik} = 0$, which follows from the fact that ${\sigma_{ X}}_{ik} = 0$ if and only if $H_{ X}(\{  w \in \Theta^+_{p-1} : w_i > 0, w_k > 0\}) = 0$.

Importantly, $\Sigma_{ X}$ has a relationship to random vectors constructed according to Corollary \ref{coro: mtxMult}.
Let $ Z = (Z_1, \ldots, Z_q)^T$ be a $q$-dimensional random vector of independent random variables such that $n \Pr( n^{-1/2}Z_j > z) \rightarrow z^{-2}$ ($j = 1, \ldots q$) and such that $n \Pr\{Z_j \leq \exp(-kn^{1/2})\} \rightarrow 0$ for any $k > 0$, $j = 1, \ldots, q$. Let $A = ( a_{1}, \ldots,  a_{q})$ be a $p \times q$ matrix with $\max_{i = 1, \ldots, p} a_{ij} \geq 0$ for all $j$. Further, assume $q \geq p$.
The results of~\S\ref{sec:  linOpsRegVar} give us that $A \circ  Z \in RV_+^p(2)$ 
with angular measure on the $L_2$ unit ball $\Theta^+_{p-1}$ of 
\begin{equation*}
  \label{eqn: angMsrMtxConstr}
  H_{A \circ  Z}(\cdot) = \sum_{j = 1}^q \|  a^{(0)}_{j} \|_2^2 \delta_{  a^{(0)}_{j} / \|  a^{(0)}_{j} \|_2} (\cdot),
\end{equation*}
and the total mass of $H_{A \circ  Z}$ is $\sum_{j = 1}^q \|  a^{(0)}_{j} \|^2_2$.
The $(i,k)$th element of $\Sigma_{A \circ  Z}$ is
\begin{equation*}
  {\sigma_{A \circ  Z}}_{ik} 
  = \int_{\Theta^+_{p-1}} w_i w_k \dd H_{A \circ  Z} ( w)
  = \sum_{j = 1}^q 
  		\left( \frac{ a_{ij}^{(0)} }{ \|  a_{j}^{(0)} \|_2 } \right) 
		\left( \frac{ a_{kj}^{(0)} }{ \|  a_{j}^{(0)} \|_2 } \right) \| 
		 a_{j}^{(0)} \|_2^2 \\
  = \sum_{j = 1}^q a_{ij}^{(0)}  a_{kj}^{(0)},
\end{equation*}
since $\alpha = 2$.  Thus $\Sigma_{A \circ  Z} = A^{(0)} (A^{(0)})^T$.



The tail pairwise dependence matrix has the additional property of being completely positive.
A matrix $S$ is completely positive if there exists a nonnegative (not necessarily square) matrix $A$ such that $S = A A^T$.

\begin{proposition}
\label{prop: completelyPositive}
If $ X$ has tail pairwise dependence matrix $\Sigma_{ X}$, there exists a $p \times q_*, q_* < \infty$, nonnegative matrix $A_{*}$ such that $\Sigma_{ X} = A_{*} A_{*}^T$.
\end{proposition}

While Proposition \ref{prop: dense} loosely says that a regularly varying random vector with any angular measure can be represented by an infinite linear combination of independent regularly varying random variables, Proposition \ref{prop: completelyPositive} says that if one restricts attention to $\Sigma_X$, this dependence structure can be represented by a finite linear combination.

As $\Sigma_X$ contains incomplete information, it is natural to ask how much information is lost by summarizing the tail dependence of $ X$ only in terms of the bivariate metrics it contains.
We investigated the extremal behavior of seven different five-dimensional random vectors which all share a common $\Sigma_X$ by calculating the measures of two different extremal sets.
We determined the probabilities that each of these random vectors takes a value in these extremal sets were similar, and the coefficients of variation of these calculated measures were 0.14 and 0.20 (see supplementary materials).

\section{Eigendecomposition and principal components}
\label{sec: eigenDecomp}
In a non-extreme setting, eigendecomposition of the covariance matrix can be motivated in two ways:  (1)  the eigenvectors form an orthonormal basis ordered in the sense of Proposition~\ref{prop: eigOrder}, and (2) the principal components defined from the eigendecomposition form an uncorrelated random vector of basis coefficients with decreasing variance.
We investigate both below.

Let $ X \in RV_+^p(2)$ such that \eqref{eq: restrictedX} holds with $\nu_{ X}$ and $H_{ X}$ the limiting and angular measures, and let $\Sigma_{ X}$ be its tail pairwise dependence matrix.
As $\Sigma_{ X}$ is positive-semidefinite, we can perform the standard eigendecomposition to obtain $\Sigma_{ X} = U D U^T$, where $D$ is the diagonal matrix with elements $\lambda_1 \geq \cdots \geq \lambda_p \geq 0$, $U = ( u_{1}, \ldots,  u_{p})$ is a $p \times p$ unitary matrix and $(\lambda_i,  u_{i})$ is an eigenvalue/eigenvector pair of $\Sigma_{ X}$ for $i = 1, \ldots, p$.
Without loss of generality, we assume that none of the vectors $ u_i$ is composed of all nonpositive elements. 
The columns of $U$ form an orthonormal basis for $\mathbb{R}^p$, and $( e_{1}, \ldots,  e_{p}) = (t( u_{1}), \ldots, t( u_{p}))$ form an orthonormal basis for $\mathbb{X}^p$. 
This basis is ordered and is most efficient in the sense of Proposition \ref{prop: eigOrder} and quadratic forms induced by $\Sigma_{ X}$.
Properties of matrix trace yield
$
  \sum_{i = 1}^p {\sigma_{ X}}_{ii} = \sum_{i = 1}^p \lambda_i.
$



If we assume that $n \Pr\{X_i \leq \exp(-kn^{1/2})\} \rightarrow 0$ for any $k > 0$, $i = 1, \ldots, p$, we can  define 
\begin{equation}
  \label{eq: pcDefn}
   V = U^T t^{-1}(X).
\end{equation}
We refer to $ V$ as the extremal principal components of $ X$.
Lemma A\ref{lem: mtxRegVar} in the appendix implies
$ V \in RV^p(2)$. 
Reversing \eqref{eq: pcDefn} and using \eqref{eq: linCombo}, we obtain an implicit definition for $ X$:
\begin{equation}
   X = U \circ  t(V) = V_1 \circ  e_1 \oplus \cdots \oplus V_p \circ  e_p.
   \label{eq: coefs}
\end{equation}
Thus the elements of $V$ are the stochastic basis coefficients when $ X$ is decomposed into the basis $ e_1,\ldots,  e_p$. 



To fully express the regular variation properties of $ V$ would require explicit knowledge of $H_{ X}$.
However, we can summarize the second-order properties of $V$ similar to before.
Define
\begin{equation*}
  {\sigma_{V}}_{ik} = \int_{\Theta_{p-1}} \omega_i \omega_k \dd H_V(\omega) 
\end{equation*}
where $\Theta_{p-1} = \{ \omega \in \mathbb{R}^p : \| w \|_2 = 1\}$, and $H_V$ is the angular measure of $V$.
Proposition \ref{prop: pcSigma} below shows that these extreme principal components have analogous properties to their non-extreme counterparts.
%


\begin{proposition}
  \label{prop: pcSigma}
  ${\sigma_{ V}}_{ii} = \lambda_i$ for all $i = 1, \ldots, p$, and ${\sigma_{ V}}_{ik} = 0$ for $i \neq k$.
\end{proposition}

The eigenvalues relate to the scale of the magnitude of the elements of $V$ as for $x > 0$, 
$$
\lim_{n \rightarrow \infty} n \Pr(n^{-1/2} | V_i |  > x) = \int_{\Theta_{p-1}} \int_{x/|\omega_i|}^\infty 2r^{-3} \dd r \dd H_V(\omega) = x^{-2} \lambda_i.
$$
Unlike for $\Sigma_X$, the fact that ${\sigma_V}_{ik} = 0$ does not imply that elements $V_i$ and $V_k$ are asymptotically independent.
Instead it implies
\begin{equation}
\label{eq: balanceSigma}
\int_{\Theta_{p-1}:\omega_1\omega_2 > 0} \omega_1 \omega_2 \dd H_V(\omega) = \int_{\Theta_{p-1}:\omega_1\omega_2 < 0} \omega_1 \omega_2 \dd H_V(\omega).
\end{equation}



\section{Completely positive decomposition}
\label{sec: compPosDecomp}
The eigendecomposition of $\Sigma_X$ can be used as an exploratory tool for tail dependence, but it does not provide a method for constructing random vectors $ X_*$ such that $\Sigma_{ X_*} = \Sigma_{ X}$.
Because $U$ contains negative entries, $\Sigma_{U \circ D^{1/2} \circ  Z} = U^{(0)} D (U^{(0)})^T \neq \Sigma_{ X}$.
However, given that there exists a $p \times q_*$ nonnegative matrix $A_{*}$  such that $\Sigma_{ X} = A_{*} A_{*}^T$, if one can find such a matrix, one can construct $ X_* = A_{*} \circ  Z_{q*}$ as in~\S\ref{sec: TPDM}.

\cite{berman2015} provides an overview of current knowledge and the status of several open problems concerning completely positive matrices.
One open problem is finding the cp-rank; that is, the minimum number of columns $q$ such that there exists a $p \times q$ matrix $A_q$ satisfying $A_q A_q^T = \Sigma$.
It is known for $p$-dimensional completely positive matrices that the cp-rank is less than $p(p+1)/2-4$ \citep{barioli2003}.
Thus the factorized matrix $A_q$ might be quite large.
\citet{Ding.etal:2005} and \citet{groetzner2016} describe algorithms to perform the factorization of $\Sigma_{ X}$ to obtain $A_q$. 
These algorithms are able to factorize matrices of moderate dimension $p \approx 50$, which is sizable for extremes studies.
Typically, one can find multiple completely positive factorizations.

\section{Applications}
\label{sec: apps}

\subsection{Estimation of $\Sigma_X$}\label{sec: estimTPDM}
To apply to data, we must estimate $\Sigma_X$.
Our estimator replaces the true angular measure with an empirical estimate.
Let $ x_t$ ($t = 1, \ldots, n_{samp}$) be independent and identically distributed vectors of observations from a random vector $ X$ whose distribution satisfies~\eqref{eq: restrictedX}, let $r_t = \|  x_t \|_2$, and $ w_t =  x_t/r_t$.
We define 
\begin{equation}\label{eqn: est}
  \hat\sigma_{{ X} ik}
  = \hat m \int_{\Theta^+_{p-1}} w_i w_k d \hat N_{ X}( w)
  = \hat m n_{exc}^{-1} \sum_{t = 1}^{n_{samp}} w_{ti} w_{tk} \mathbb{I}(r_t > r_0), 
\end{equation}
where $r_0$ is some high threshold for the radial components, $n_{exc} = \sum_{t = 1}^{n_{samp}} \mathbb{I}(r_t > r_0)$, $N_{ X}(\cdot) = m^{-1} H_{ X}(\cdot)$, and $\hat m$ is an estimate of $H_{ X}(\Theta^+_{p-1})$.
The estimator in \eqref{eqn: est} is the same as that given by \cite{larsson2012} in the bivariate case. 
Because we preprocess the data to have a common unit scale in the two applications below, $m = p$ and does not need to be estimated.  
When the data are not preprocessed to have a common scale an empirical estimator is $\hat m = \frac{r_0^2}{n_{samp}} \sum_{t = 1}^{n_{samp}} \mathbb{I}(r_t > r_0)$.
For principal component analysis, preprocessing the data to have unit scale is analogous to performing the eigendecomposition on the correlation rather than the covariance matrix in the non-extreme setting.

As $\hat\Sigma_{ X}=n_{exc}^{-1} \hat m \hat W^T \hat W$ where $\hat W$ is the matrix whose rows are the vectors $ w_t$ for which $r_t>r_0$, the estimate $\hat\Sigma_{ X}$ is positive semidefinite and completely positive.
The observed matrix $\hat W$ provides a perhaps inefficient completely positive factorization of $\hat \Sigma_X$:  $\hat A_{*} = n_{exc}^{-1/2} \hat m^{1/2} \hat W^T$.
In applications we will find a completely positive factorization with fewer columns thereby reducing the dimension compared to the inefficient factorization.
In addition to the two examples below, we perform a simulation study which shows that eigenvectors of $\hat \Sigma_X$ retain the interpretable dependence information of the eigenvectors of $\Sigma_X$.
Furthermore, eigenvalues of $\hat \Sigma_X$ have similar behavior to eigenvalues of estimated covariance matrices  (see supplementary materials).

\subsection{Extreme precipitation in Switzerland}
We analyze precipitation measurements from $p = 44$ stations located near Z\"urich, Switzerland, obtained from M\'et\'eoSuisse.
Measurements are daily precipitation amounts (in mm) for June, July, and August for the years 1962--2012, yielding $n_{samp} = 4691$ observations. 
The same data were analyzed by \cite{thibaud2015} and were found to be asymptotically dependent.

The data appear to be heavy-tailed (location-wise estimates for $\alpha$ have a median of $7.17$); however, our framework requires that $\alpha = 2$.
Since extremal dependence is often described assuming conditions on the univariate marginals, it is common practice to incorporate marginal transformations within an extreme value analysis \citep[\S\S6.5.3, 9.2.3]{resnick2007}. 
Letting $ X_t^{(orig)}$  be the random vector representing the precipitation measurements on day $t$, a marginal transformation $g$ can be applied so that $ X_t = g(  X^{(orig)}_t) = \{g_1(X^{(orig)}_{t1}), \ldots, g_p(X^{(orig)}_{tp})\}^T$ has the desired marginal properties given in~\eqref{eq: restrictedX}.

We perform a nonparametric marginal transformation. 
We define $g_i(x) = \{-\log \hat F_i(x)\}^{-1/2}$,
so that $\Pr(X_{ti} \leq x) \approx \exp(-x^{-2})$. 
With this transformation, the scale of $X_{ti}$ is 1 for $i = 1, \ldots, p$. 
We employ a linearly-interpolated empirical cumulative distribution function for $\hat F_i$. 
We let $r_0=33.4$ which corresponds to the empirical $0.95$ quantile, yielding $n_{exc}=235$ large observations to estimate $\Sigma_{ X}$.


We perform the eigendecomposition to find $\hat U \hat D \hat U^T = \hat \Sigma_{ X}$.
Figure \ref{fig: precipEvectors} shows $\hat { e}_{1}=t(\hat { u}_{1}), \ldots, \hat { e}_{5}=t(\hat { u}_{5})$, the first five eigenvectors of $\hat \Sigma_{ X}$ in $\mathbb{X}^p$, plotted according to each station's location. 
As the zero element of $\mathbb{X}^p$ is $\log 2$, we will refer to values less than $\log 2$ as ``negative".
Although our analysis does not use location information, the leading eigenvectors clearly large-scale spatial behavior whose resolution increases with order.
The first eigenvector has values ranging only from $0.332$ to $0.337$,
implying that the leading eigenvector accounts for the overall magnitude of the precipitation event. 
The second and third eigenvectors show basically linear trends; the second eigenvector decreases from positive values in the northwest to negative values in the southeast, and the third has decreasing values southwest to northeast.  
The fourth eigenvector shows behavior which is roughly quadratic, with the lowest values in the center of the region and higher values in the north and south, and, to a lesser extent, west.
The fifth eigenvector shows saddle-like behavior with moderate values in the center, low values to the northwest and southeast, and high values to the southwest and northeast.

Also shown in Figure~\ref{fig: precipEvectors} is a scree plot of the eigenvalues.  
Because of preprocessing $\sum_{i = 1}^p \lambda_i = p = 44$.
The first five eigenvalues are $30.15$, $2.62$, $1.39$, $0.90$, $0.79$, and the scree plot shows that the magnitudes of the eigenvalues become quite small after the first few values.


\begin{figure}[t]
  \begin{center}
    \includegraphics[width = .32\textwidth]{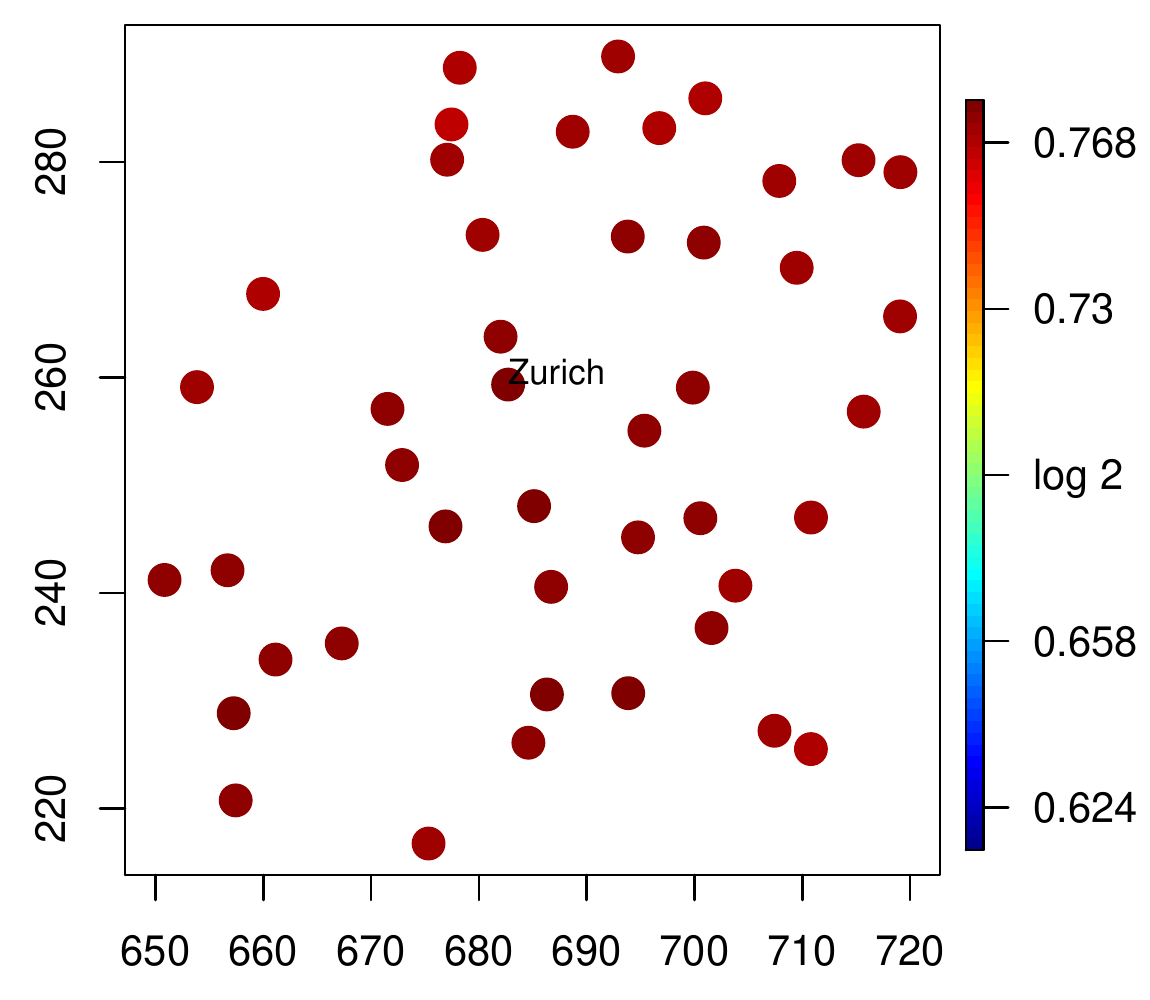}
    \includegraphics[width = .32\textwidth]{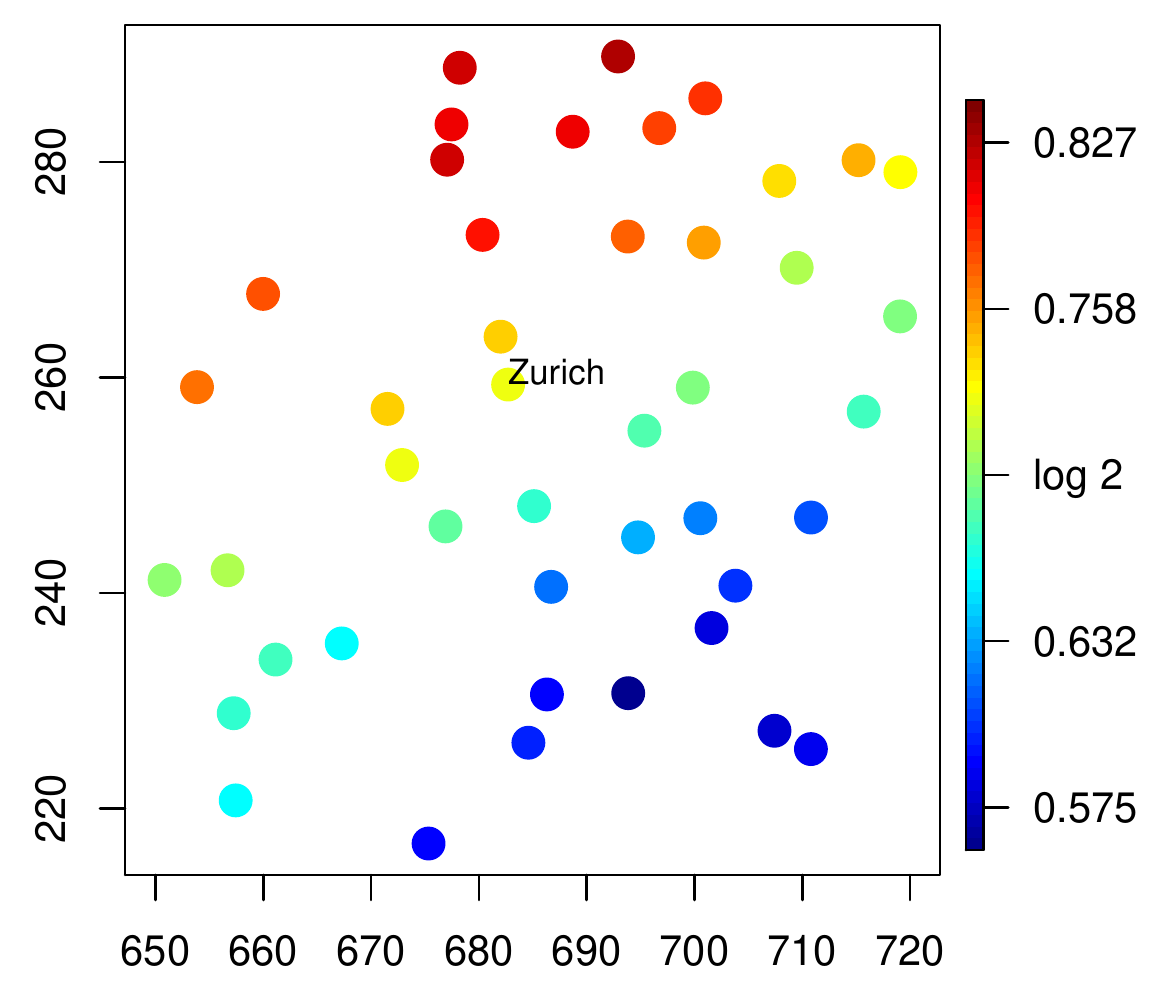}
    \includegraphics[width = .32\textwidth]{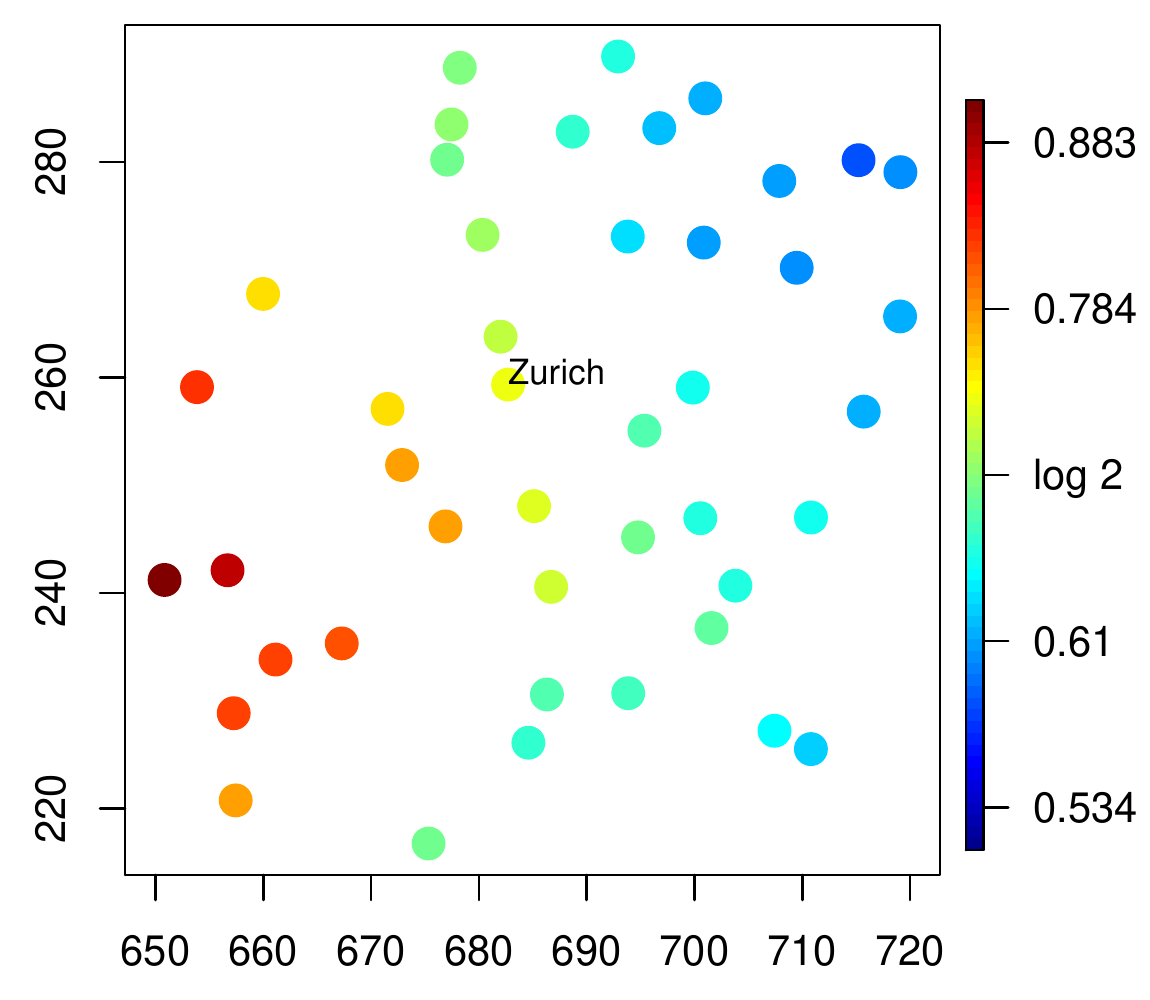}\\
    \includegraphics[width = .32\textwidth]{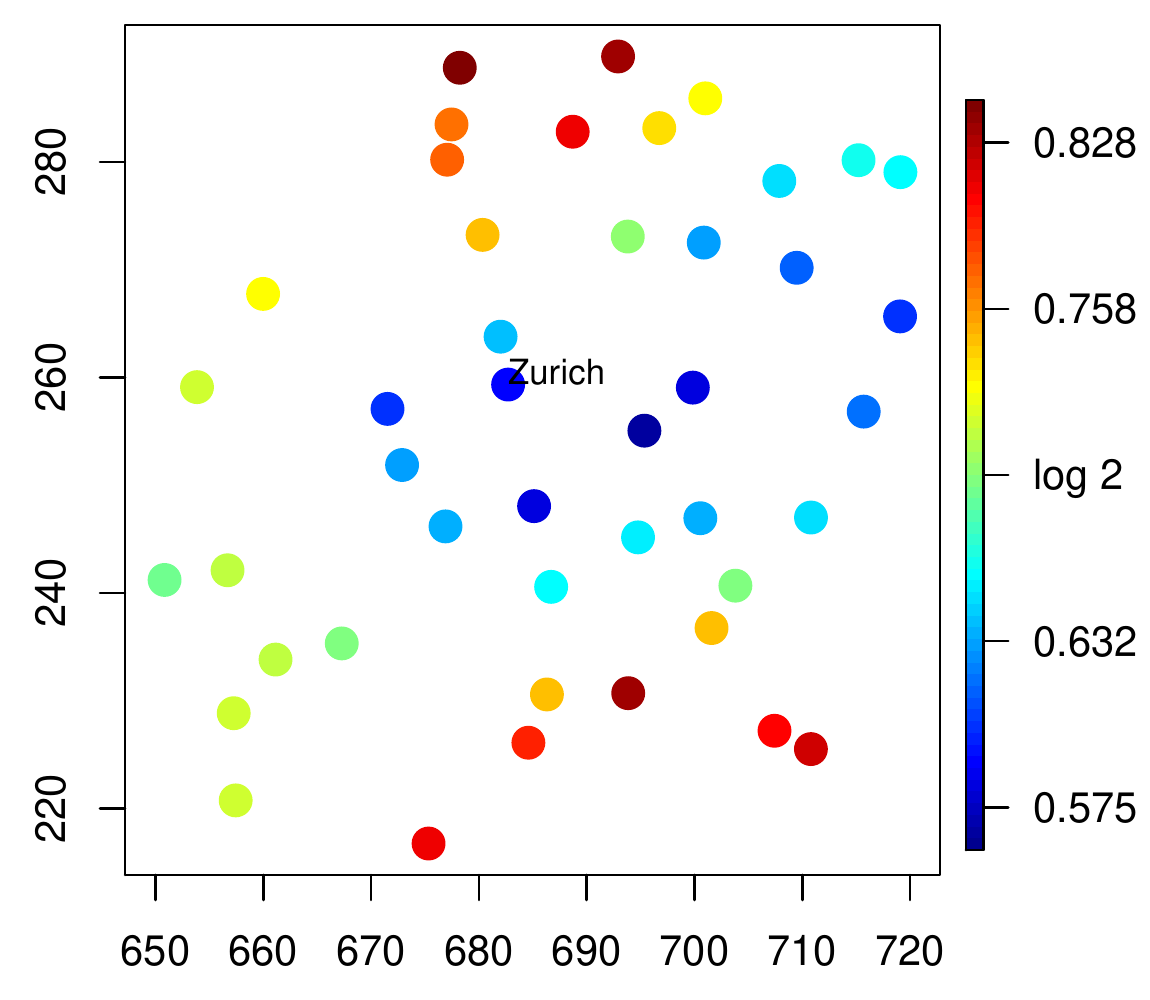}
    \includegraphics[width = .32\textwidth]{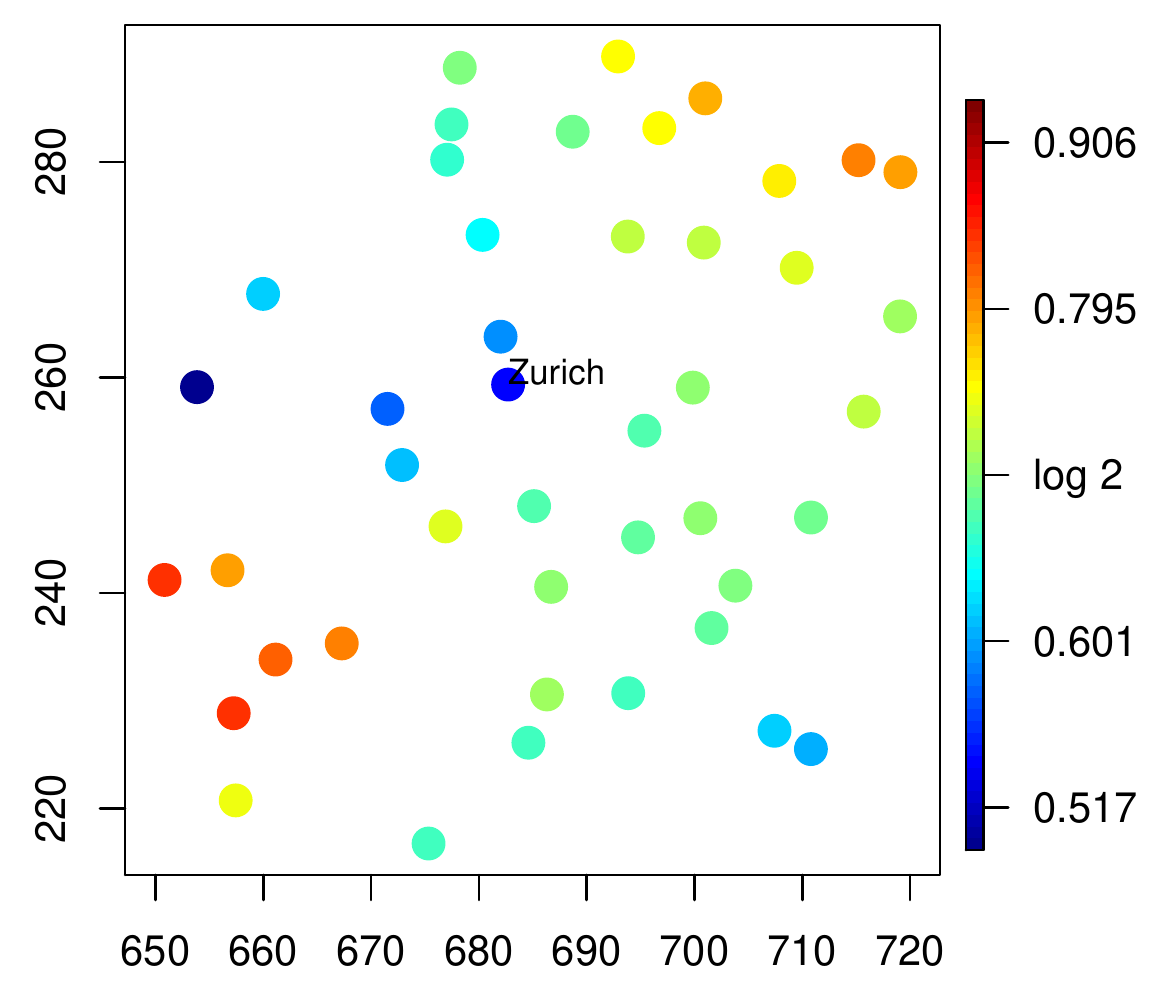}
    \includegraphics[width = .32\textwidth]{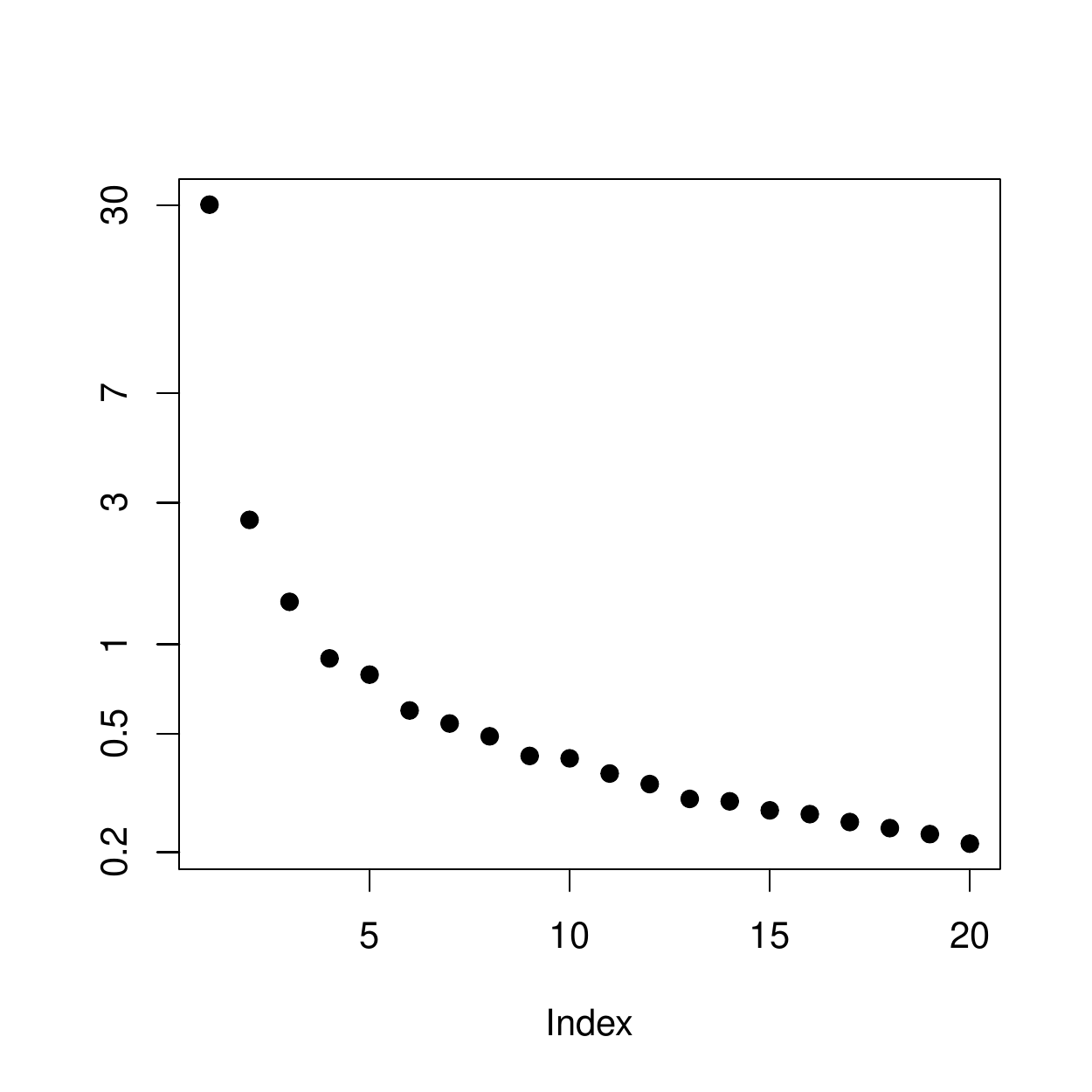}
  \end{center}
  \caption{Plots of the first five eigenvectors in $\mathbb{X}^p$. Color scales are balanced so that $\log 2=t(0)$ corresponds to the same color in each figure. However, note that each plot has its own scale.  Axis labels are in km. The lower right panel is a scree plot of eigenvalues 
   on log scale.}
  \label{fig:  precipEvectors}
\end{figure}

\begin{figure}[t]
  \begin{center}
    \includegraphics[width = .32\textwidth]{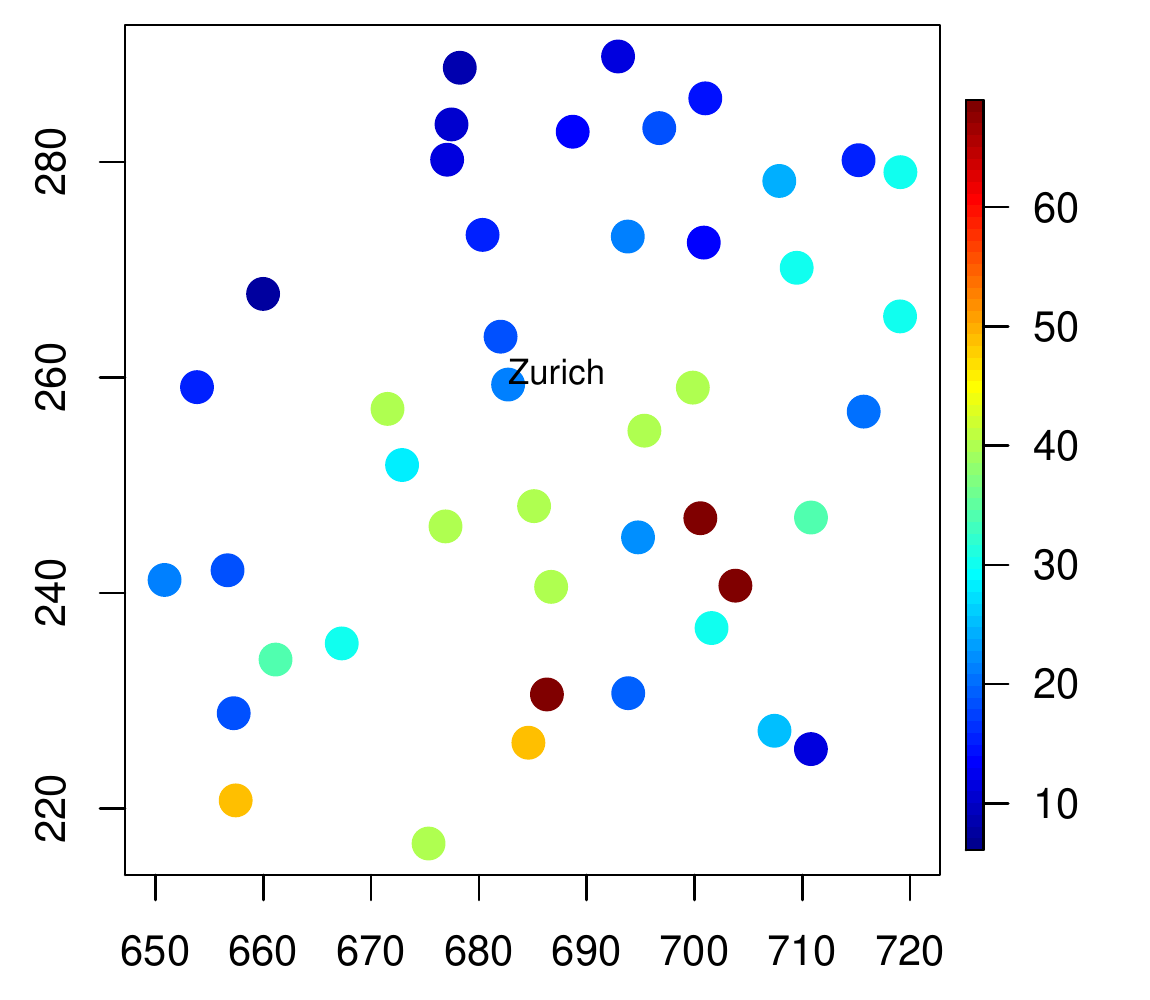}
    \includegraphics[width = .32\textwidth]{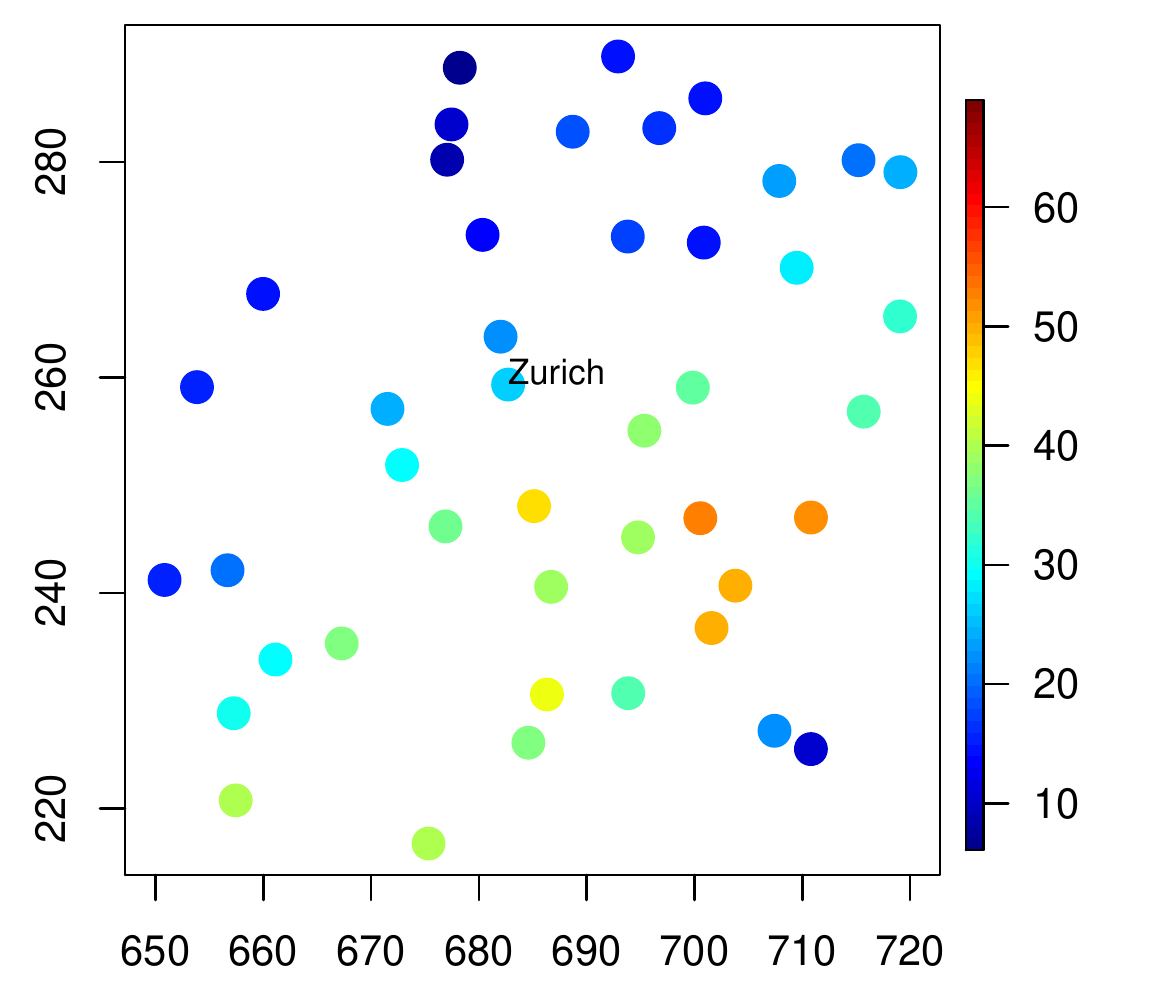}
    \includegraphics[width = .32\textwidth]{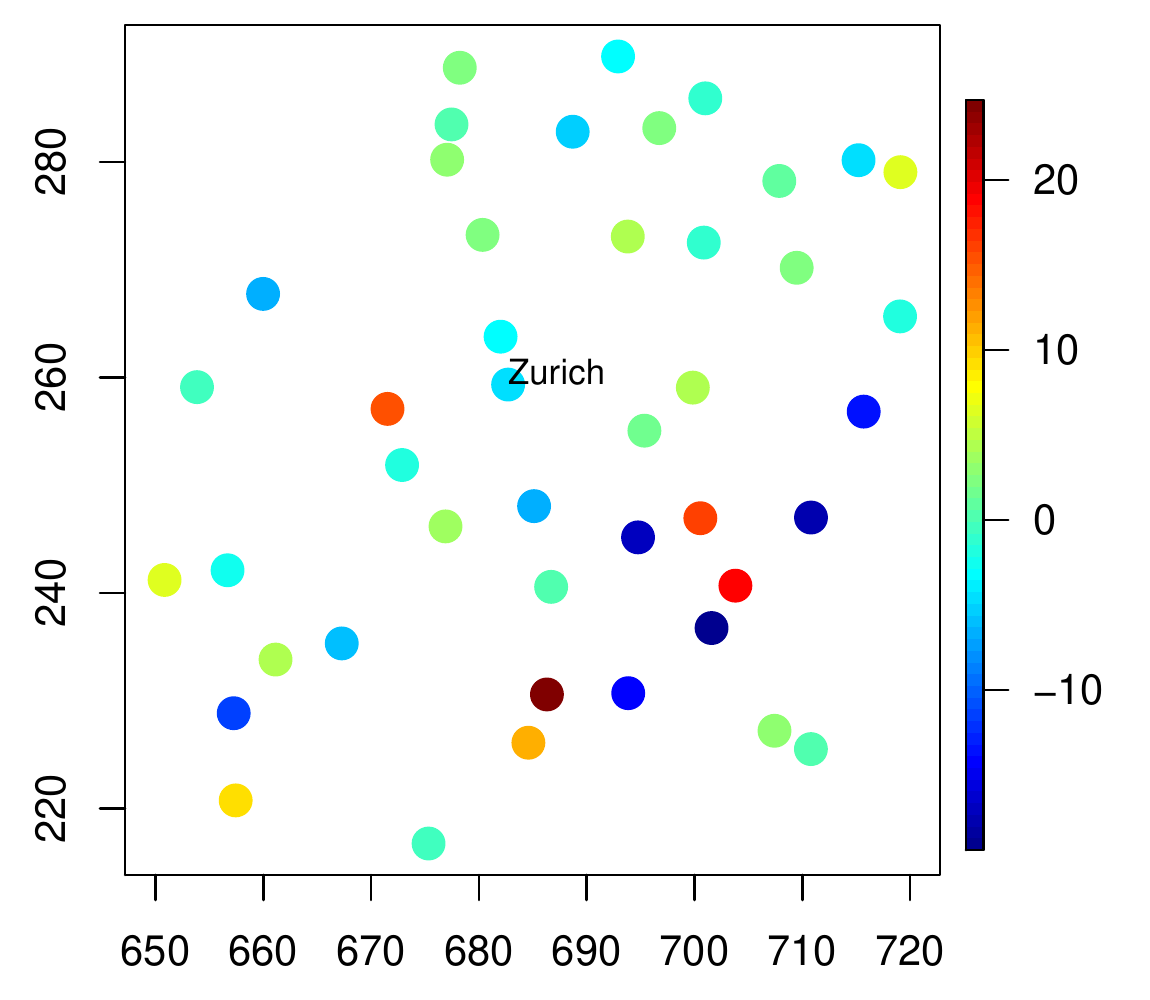}
  \end{center}
  \caption{Left:  Plot of the transformed observations from the third-largest event.  Center:  Reconstruction of event using only the ten largest basis functions.  Right:  Difference between observations and reconstruction.}
  \label{fig: recnstr}
\end{figure}

We find the sample principal components by letting $ v_t = \hat U^T t^{-1} (x_t)$.
Figure \ref{fig: recnstr} illustrates a partial basis reconstruction of an event.
The left panel displays the transformed observations from the day with the third-largest value of $r_t$ in our record.
For reference, $\Pr(X_{ti} \leq 9.97) = 0.99$ and $\Pr(X_{ti} \leq 31.6) = 0.999$, so values at all locations on this day are large.
Observations from this day are generally increasing from the northwest to the southeast, until one reaches the extreme southeast corner of the study region.
Letting $t^*$ denote this particular day, we could represent $ x_{t^*}$ exactly as a linear combination of the basis
\begin{equation}
  \label{eq: recnstr}
   x_{t^*} = \hat U \circ t(v_{t^*}) = v_{t^*1} \circ  e_{1} \oplus \cdots \oplus v_{t^*44} \circ  e_{44}
\end{equation}
from (\ref{eq: linCombo}).
The center panel of Figure \ref{fig: recnstr} shows (\ref{eq: recnstr}) truncated after the first ten terms in the transformed linear combination.
We see that the general nature of the event is largely reconstructed from these leading basis vectors, with the increase from northwest to southeast, and the lower values in the extreme southeast corner.
As expected, some of the fine-scale behavior is not reproduced using only the leading ten basis vectors, as is shown in the difference plot in the right panel.
It is also interesting to look at the observed principal components $\{v_{t^*1}, \ldots, v_{t^*5}\} = (187.5, -59.6,  -2.9, -27.4,  4.9)$. 
The large value of the first principal component when paired with $e_1$  gives large values to all locations.
The large negative value of the second principal component when paired with the $e_2$ largely contributes the northwest to southeast increase shown in the reconstruction.

We obtain a completely positive factorization of $\hat \Sigma_{ X}$ using the method of \cite{groetzner2016}. 
We find $\hat A_*$ with only $51$ columns such that $\hat \Sigma_{ X} = \hat A_{*} \hat A_{*}^T$. 
We define $  X_* = \hat A_{*} \circ   Z$ where $Z_j$ are independent and $\Pr( Z_j \leq z) = \exp(-z^{-2})$ ($j = 1, \ldots, 51$).
Although $  X_*$ shares the same tail pairwise dependence matrix as the one estimated for $X$, they do not share the same angular measure.

Because the angular measure of $  X_*$ is discrete, probabilities of extreme events (i.e. $\Pr(  X_* \in C)$ for some set of interest $C$) can be calculated easily.
We define an initial risk region on the scale of the original data:  $C^{(orig)} = \{  x \in \mathbb{R}^{44} \mid   x \in [30, \infty]^{44}\}$, and then calculate $\Pr(  X_* \in C)$ where $C = g(C^{(orig)})$.
Letting $\hat a_{ij}$ be the elements of $\hat A_{*}$, $\hat \Pr(  X_* \in C) = \sum_{j = 1}^{51} \min_{i = 1, \ldots p} {\hat a_{ij}^2}/{{g_i^2(30)}}$, and our probability estimate is $4.8 \times 10^{-4}$.
An empirical estimate of $\Pr(  X^{(orig)} \in C^{(orig)})$ is 2/4691 = $4.3 \times 10^{-4}$.
Uncertainty in the probability estimate arises from two sources:  uncertainty in the estimate $\hat \Sigma_X$ and the non-uniqueness of the completely-positive factorization.  
A bootstrap based confidence interval could include both sources of uncertainty, but this is infeasible as there are not readily-available methods for repeatedly obtaining completely-positive factorizations for $p = 44$.
Interestingly, using the $\hat A_{*}$ with 235 columns arising from the inefficient factorization (see \S\ref{sec: estimTPDM}) yields an estimate of $1.7 \times 10^{-3}$.

\subsection{Extreme losses for financial data}

Our data are the `value-averaged' daily returns of 30 industry portfolios compiled and posted as part of the Kenneth French Data Library. 
We analyze data for 1950--2015, yielding $n_{samp} = 16694$ observations.
We transform the data to perform our analysis. 
Let $ x_t^{(orig)}$ denote the vector of returns for day $t$.
Then, let $ x^{(temp)}_{ti} = t\big[ \{\max(- x_{ti}^{(orig)}, 0)\}_{i = 1, \ldots, p} \big]$, negating the returns since we are interested in extreme losses, setting negative values (gains) to zero, and then applying our transform $t$ to bound the values away from $0$ thereby meeting the lower-tail requirements for $ X$. 
Importantly, this transformation leaves the magnitudes of the large losses essentially unchanged.
We use the Hill estimator \citep{hill1975} at the empirical $0.99$ quantile to obtain estimates $\hat \alpha_i$ for each marginal.
The estimates $\hat \alpha_i$ range from $2.75$ to $3.81$ with the heaviest tails belonging to the categories Finance (Fin), Steel, Textiles (Txtls), Coal, and Telecom (Telcm).
Estimated scales of the variables also varied widely and eigenvectors of the tail pairwise dependence matrix of the unscaled data were dominated by the variable scales.
Consequently, we let $x_{ti} = \hat c_i^{-1/2} (x^{(temp)}_{ti})^{\hat \alpha_i/2}$  ($i = 1, \ldots, p$) where $\hat c_i$ is the $i$th component's scale estimate, allowing us to assume a tail index of $\alpha = 2$ and a scale of 1 for all marginals.  
$\hat \Sigma$ is then estimated as before, employing data whose radial components exceed the 0.99 quantile.

\begin{figure}[t]
  \begin{center}
    \includegraphics[scale = .65]{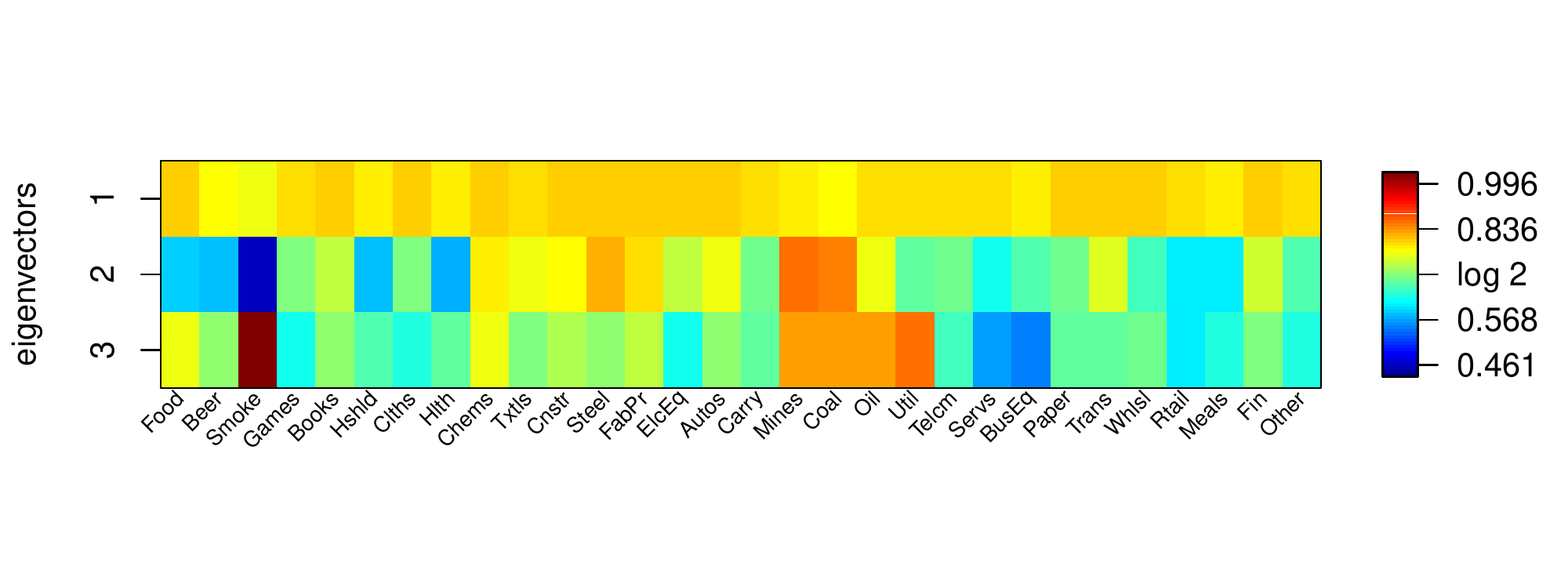}\\[-7ex]
    \includegraphics[scale = .6]{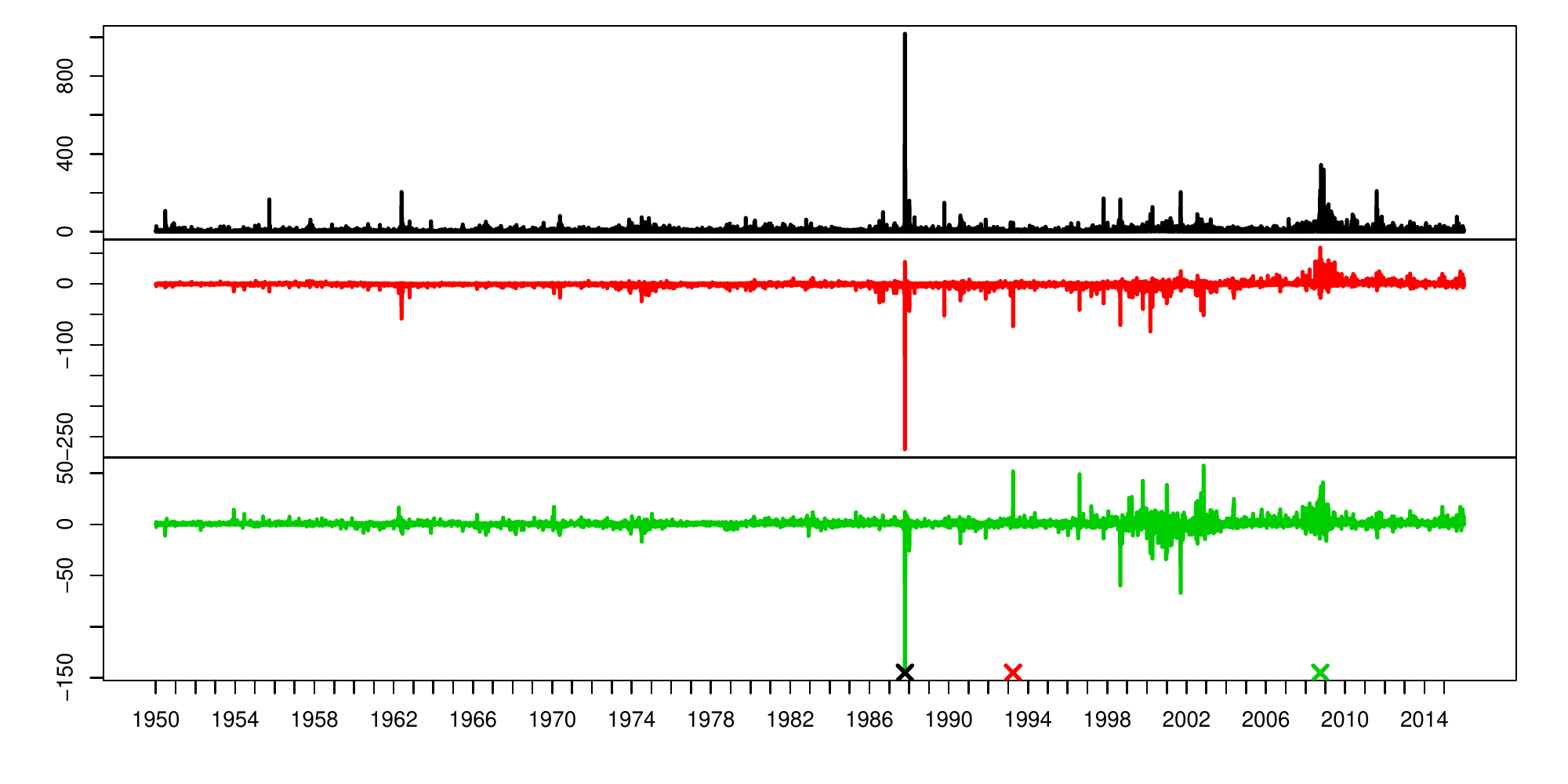}\\[-3ex]
    \includegraphics[scale = .57]{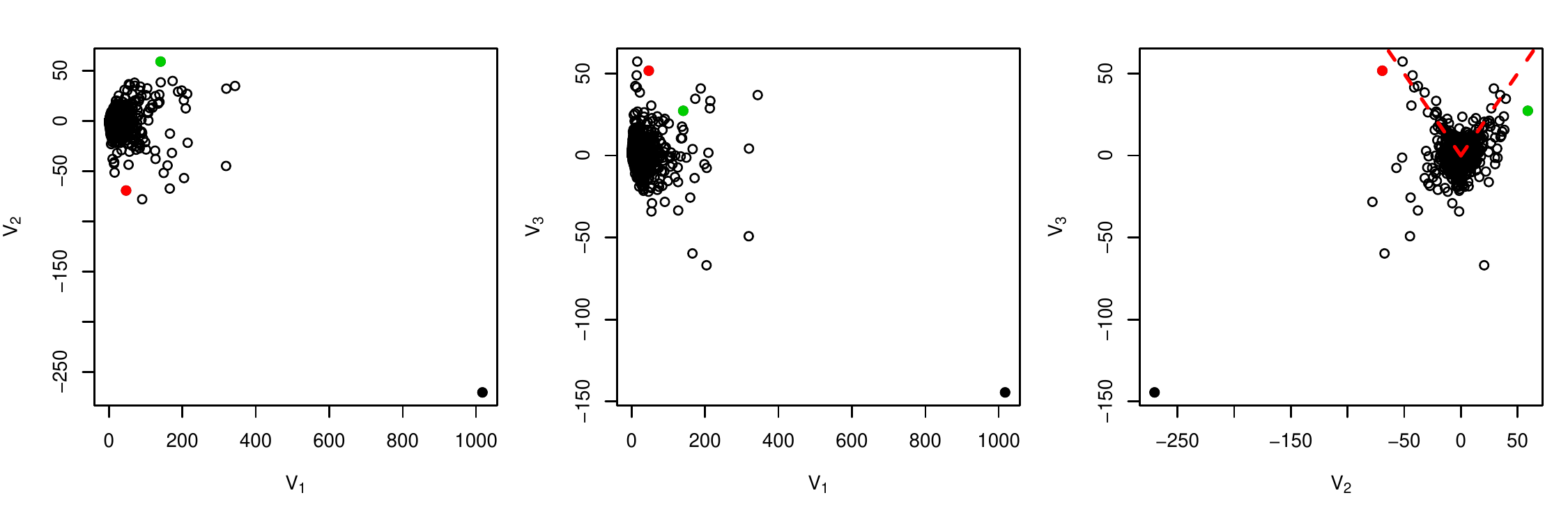}
  \end{center}
  \vspace*{-0.5cm}
  \caption{Leading three eigenvectors for the financial data (top), time series of principal components (middle), and scattterplots of pairs of principal components (bottom). Note that $\log 2$ should be interpreted as the 0 element in the top panel showing eigenvector values.}
  \label{fig:  finData}
\end{figure}

We perform an eigendecomposition of $\hat \Sigma_{ X}$ and $\lambda_i/\sum_{j=1}^p \lambda_j$ is 0.680, 0.052, and 0.036 for $i = 1, 2, 3$.
The top panel of Figure \ref{fig: finData} shows the values of the first three eigenvectors $e_i$,  values less than $\log 2$ are again referred to as ``negative".
Eigenvector $e_1$ gives an overall magnitude, $e_2$ contrasts ``heavy" and ``non-heavy" industries with Mines, Coal, and Steel having the largest positive values versus Smoke, Food, Beer, Consumer Goods (Hshld), and Healthcare (Hlth) with the largest negative values.
The third eigenvector $e_3$ contrasts Smoke, and several heavy industries (Mines, Coal, Oil, and Utilities (Util)) versus ``office/executive" sectors Services (Servs) and Business Equipment (BusEq).
The middle panel shows the time series of the first three principal components $v_{t i}$ ($i = 1, 2, 3$); note the different scales for the time series plots.
The time series clearly shows ``Black Monday", October 19, 1987, marked with a black `X'.
The large positive value of the first coefficient arises from the large losses across the market, while the large negative value of the second coefficient helps to mitigate this affect for Mines and Coal.
The median of log-returns across sectors on Black Monday was $-17.61$, but Mines and Coal were not as severely affected with log returns of $-9.87$ and $-11.58$ respectively. 
Also, it is interesting that the third coefficient experiences volatility in 1999--2002 time period and the first and second do not.
This likely reflects that this time period corresponding to the American tech bubble was troublesome for Business Equipment and Services whose definitions include many computer-related items.
The bottom panel shows the pairwise scatterplots of the first three principal components.
Clearly these principal components are not asymptotically independent as large values do not occur on the axes.
The right plot of principal component three versus two shows extreme behavior in two directions:  $(-1/\sqrt 2, 1/\sqrt 2)$ and $(1/\sqrt 2, 1/\sqrt 2)$ which are marked.
The two points indicated in red and green are the points with the largest values (other than Black Monday which is marked in black) when projected in these directions.
The red point corresponds to April 2, 1993, known as ``Marlboro Friday", and the green point corresponds to October 2, 2008 which occurred during the financial crisis of 2008, and these dates are indicated in the time series plots.
The change in sign of the second principal component implies the market behaves very differently in these two directions.
Marlboro Friday had large losses across the market (median log-return of $-1.8$), particularly large losses for Food and Smoke ($-7.9$ and $-12.8$ respectively), but moderate gains for Mines and Coal ($0.4$ and $0.5$).  
October 2, 2008 shows very different behavior with a median log return of $-4.7$ across all sectors, but particularly big losses for Mines and Coal ($-12.8$ and $-13.7$ respectively).
The fact that there are a number of large values in these directions indicates this type of behavior is not unique to just these two days.



\section{Discussion} 
\label{sec: disc}

This work novelly applies linear algebra constructs within the context of extremes, achieved by defining a vector space for the positive orthant via a transform $t$.
Importantly, the vector space yields the idea of basis for the space in which our regularly varying random vectors take values, and the specific transform preserves regular variation.


Extremal dependence is summarized by the tail pairwise dependence matrix $\Sigma_X$.
Although we are not the first to summarize tail dependence via a collection of summary metrics \citep[e.g.,][]{strokorb2013}, we believe the similarities between the tail pairwise dependence matrix and the covariance matrix are appealing.
The eigendecomposition of $\Sigma_X$ allows interpretation of dependence via the eigenbasis, very similar to traditional principal component analysis.
The fact that the eigendecomposition does not lead to a construction method for a random vector with the desired tail pairwise dependence matrix is overcome by the completely positive decomposition. 


Importantly, we do not attempt to characterize the $p-1$ dimensional angular measure, as we think this is very difficult when $p$ is large. 
Much of traditional multivariate analysis is based on second-moment characterizations, and it is often found that having all pairwise summaries provides adequate information for the dependence to be understood, for models to be constructed and used to estimate quantities of interest, and ultimately, for decisions to be made. 

Beyond the decompositions described in this work, the connection between extremes and linear algebra warrants further investigation.
Given the prevalence of linear manipulations in multivariate analysis, spatial statistics, and time series, these linear algebra connections could lead to further models and methodological development for extremes.

\section*{Acknowledgment}  D. Cooley and E. Thibaud were both partially supported by NSF DMS-1243102.  We thank Patrick Groetzner and Mirjam D\"ur for their help with completely positive matrix factorizations, and Piotr Kokoszka for helpful discussion. The financial data were obtained from http://mba.tuck.dartmouth.edu/pages/faculty/ken.french/data\_library.html.   We thank the editors and reviewers for helpful comments.



\begin{appendix}
\section{Appendix}

\subsection{Lemmas}
\label{app:  lemmaProof}


Before considering `transformed-linear' operations applied to regularly varying random vectors, the two following lemmas show the transform $t$ and its inverse preserve regular variation between $RV_+^p(\alpha)$ and $RV^p(\alpha)$.

\begin{lemma}\label{lem:tm1XisRV}
Let $X \in RV_+^p(\alpha)$ such that $n \Pr(b_n^{-1} X \in \cdot)\stackrel{v}{\rightarrow} \nu_{X} (\cdot)$ and condition~\eqref{eq:lowertailcond} holds. 
Then $t^{-1}( X) \in RV^p(\alpha)$, and 
\begin{equation}\label{eq:convtm1X}
n \Pr\{ b_n^{-1} t^{-1}(X) \in \cdot \} \stackrel{v}{\rightarrow} \nu_{t^{-1}( X)}(\cdot) = \nu_{ X}(\cdot \cap \bar{\mathbb{X}}^p).
\end{equation}
\end{lemma}

Intuitively, because $t^{-1}$ negligibly affects large positive values, if $C \subset \bar{\mathbb{X}}^p\setminus\{ 0\}$, then $\nu_{t^{-1}( X)}(C) = \nu_{ X}(C)$.
The lower-tail condition \eqref{eq:lowertailcond} on the marginals of $ X$ implies that if $C \subset ({\bar{\mathbb{X}}^p})^c \setminus 0$ then $n\Pr\{b_n^{-1} t^{-1}(X) \in C \}\rightarrow 0$, i.e., $\nu_{t^{-1}( X)}$ has no mass outside the nonnegative orthant.

\begin{lemma}\label{lem:tYisRV}
Assume $ Y \in RV^p(\alpha)$, $n \Pr(b_n^{-1}  Y \in \cdot ) \stackrel{v}{\rightarrow} \nu_{ Y} (\cdot)$. 
Then $t( Y) \in RV_+^p(\alpha)$, and 
\begin{equation*}\label{eq:convtY}
n \Pr\{ b_n^{-1} t( Y) \in \cdot \} \stackrel{v}{\rightarrow} \nu_{t( Y)}(\cdot) =  \nu_{ Y}(\{  y \in \bar{\mathbb{R}}^p :  y^{(0)} \in \cdot \}),
\end{equation*}
where $ y^{(0)} = \max( y,  0)$ applied componentwise.
\end{lemma}

Intuitively Lemma~A\ref{lem:tYisRV} says that the mass of $\nu_{ Y}$ outside the nonnegative orthant is projected on the boundary of $\mathbb{R}_+^p$ when applying the transform $t$ to $ Y$.
If $Y = t^{-1}(X)$ where $X$ meets the lower-tail condition~\eqref{eq:lowertailcond} then $\nu_{Y}$ only has mass on $\bar{\mathbb{X}}^p$, and consequently, for any $C\subset\bar{\mathbb{X}}^p\setminus\{ 0\}$, $\nu_{t( Y)}(C)=\nu_{ Y}(\{  y \in \bar{\mathbb{R}}^p :  y^{(0)} \in C \}) = \nu_{ Y}(C)$.

Before proving Lemma A\ref{lem:tm1XisRV}, we first prove the convergence in (\ref{eq:convtm1X}) for rectangular sets.
Being a componentwise function, $t$ and its inverse $t^{-1}$ apply nicely to rectangular sets.

\begin{lemma}
\label{lem:rectangleConv}
Assume $ X$ is as in Lemma~A\ref{lem:tm1XisRV}. 
Let $[ l,  u]$ be a rectangle in $\bar{\mathbb{R}}^p\setminus\{ 0\}$ where $l_i \neq 0, u_i \neq 0$ for all $i = 1, \ldots, p$.  Then
\begin{equation*}\label{eq:lemmarect}
n \Pr\{ b_n^{-1} t^{-1}( X) \in [ l,  u]\} \rightarrow
\begin{cases}
  0 ,&  u \ngtr  0, \\
  \nu_{ X}( [\max( l,  0),  u]), & u >  0.
  \end{cases}
\end{equation*}
\end{lemma}
{\em Proof of Lemma A\ref{lem:rectangleConv}:}  Note that
\begin{equation*}
n \Pr\{ b_n^{-1} t^{-1}( X) \in [ l,  u]\} = n \Pr\{ X \in t(b_n [ l,  u]) \}.
\end{equation*}
First, suppose $u_i < 0$ for some $i \in \{1,\ldots,p\}$.   Then
\begin{align*}
  n \Pr \{ X \in t(b_n [ l,  u]) \} 
  &\leq n \Pr\{ X_i \leq t(b_n u_i) \} \\
  &\sim n \Pr\{ X_i \leq \exp(-k b_n u_i)\} \rightarrow 0, \mbox{ by assumption.}
\end{align*}
Second, for $ u >  0$,
\begin{align*}
  n \Pr\{ X \in t(b_n [ l,  u]) \}
  &\sim
  n \Pr 
  \left\{  X \in \left[  
    \left( 
      \begin{array}{c l} 
        \exp(b_n l_i) & \mbox{ if } l_i < 0\\
        b_n l_i & \mbox{ if } l_i > 0
      \end{array}
    \right)_{i = 1, \ldots , p},
    (b_n u_i)_{i = 1, \ldots, p}
  \right] \right\}\\
  &=
  n \Pr\{ X \in [\max(b_n  l,  0), b_n  u]\}\\
  &\quad -
  n \Pr 
  \left\{  X \in \left[  
    \max(b_n  l,  0), 
    \left( 
      \begin{array}{c l} 
        \exp(b_n l_i) & \mbox{ if } l_i < 0\\
        b_n u_i & \mbox{ if } l_i > 0
      \end{array}
    \right)_{i = 1, \ldots , p}
  \right] \right\} \\
  &\rightarrow
  \nu_{ X}( [ \max( l,  0),  u] ) - 0.
\end{align*}

Remark: The reason Lemma~A\ref{lem:rectangleConv} excludes rectangles with vertices $l_i=0$ or $u_i = 0$ is that these rectangles would have edges at $t(0) = \log 2$ after applying the transform, and therefore would not scale radially as $n$ increases, and thus their limiting measure is unknown. \smallskip

\noindent {\em Proof of Lemma~A\ref{lem:tm1XisRV}:}  Our proof of regular variation for $t^{-1}( X)$ is similar to the proof of Lemma 6.1 in \citet{resnick2007} or the proof of \citet[][\S2]{basrak2002}: first we show that the sequence of measures $[n \Pr\{ {t^{-1}( X)}/{b_n} \in \cdot\}]_{n}$ is tight in $\bar{\mathbb{R}}^p\setminus\{ 0\}$, and second we show that all subsequential limits agree on a $\pi$-system that generates the Borel $\sigma$-algebra.

First, any bounded Borel set $B\subset \bar{\mathbb{R}}^p\setminus\{ 0\}$ is contained in the union of a finite number of rectangles $R_1=[ l_1, u_1],\cdots,R_K=[ l_K, u_K]$ in $\bar{\mathbb{R}}^p\setminus\{ 0\}$ and with no edges on the axes, so that
\begin{equation*}
\sup_{n\geq 1} n\Pr\{ b_n^{-1} t^{-1}( X) \in B \} \leq \sup_{n\geq 1} \sum_{k=1}^K n\Pr\{ b_n^{-1} t^{-1}(X) \in R_k \} < \infty
\end{equation*}
by Lemma~A\ref{lem:rectangleConv}. Hence the sequence of measure $n \Pr\{ b_n^{-1} t^{-1}(X) \in \cdot \}$ is tight in $\bar{\mathbb{R}}^p\setminus\{ 0\}$.
Second, the convergence in Lemma~A\ref{lem:rectangleConv} occurs on the $\pi$-system of rectangles with no edges on the axis and thus occurs on the $\sigma$-algebra generated by these rectangles, which coincides with the Borel algebra on $\bar{\mathbb{R}}^p\setminus\{ 0\}$ (rectangles with no edges on the axes are sufficient to construct any open rectangle of $\bar{\mathbb{R}}^p\setminus\{ 0\}$ using countable union).\smallskip

\noindent {\em Proof of Lemma~A\ref{lem:tYisRV}:} The proof is similar to that of Lemma~A\ref{lem:tm1XisRV}. Consider rectangles $[l, u] \subset  \bar{\mathbb{X}}^p\setminus\{0\}$, $ l\geq 0$, $ u > 0$. Then 
\begin{align*}
  n \Pr\{b_n^{-1} t(Y) \in [ l,  u] \}
  &=
  n \Pr\{ Y \in [t^{-1}(b_n  l), t^{-1}(b_n  u)] \}\\
  &\sim
  n \Pr 
  \left\{  Y \in \left[  
    \left( 
      \begin{array}{c l} 
        -\infty & \mbox{ if } l_i = 0\\
        b_n l_i & \mbox{ if } l_i > 0
      \end{array}
    \right)_{i = 1, \ldots , p},
    (b_n u_i)_{i = 1, \ldots, p}
  \right] \right\}\\
  &=
  n \Pr 
  \left\{ b_n^{-1} Y \in \left[  
    \left( 
      \begin{array}{c l} 
        -\infty & \mbox{ if } l_i = 0\\
        l_i & \mbox{ if } l_i > 0
      \end{array}
    \right)_{i = 1, \ldots , p},
    (u_i)_{i = 1, \ldots, p}
  \right] \right\}\\
  &\rightarrow
  \nu_{ Y}\left( \left[  
    \left( 
      \begin{array}{c l} 
        -\infty & \mbox{ if } l_i = 0\\
        l_i & \mbox{ if } l_i > 0
      \end{array}
    \right)_{i = 1, \ldots , p},
    (u_i)_{i = 1, \ldots, p}
  \right] \right) \\
  &=
  \nu_{ Y}(\{ y \in \bar{\mathbb{R}}^p :  y^{(0)} \in [ l, u]  \}),
\end{align*}
for any rectangle that is a continuity set of the limiting measure. We conclude the proof by noting that the sets of rectangles $[ l, u]$ with $ l\geq 0$ and $ u > 0$ form a $\pi$-system which generates the Borel $\sigma$-algebra of $\mathbb{R}_+^p$.

\subsection{Proofs of propositions and corollaries}

\noindent {\em Proof of Proposition~\ref{prop: sum}:}  From Lemma~A\ref{lem:tm1XisRV} the random vectors $t^{-1}(  X_1), t^{-1}(  X_2) \in RV^p(\alpha)$ with respective measures $\nu_{t^{-1}(  X_1)}(\cdot) =\nu_{ X_1}(\cdot \cap \bar{\mathbb{X}}^p)$ and $\nu_{t^{-1}(  X_2)}(\cdot) =\nu_{ X_2}(\cdot \cap \bar{\mathbb{X}}^p)$ when normalized by $\{b_n\}$. 
Proposition~7.4 of \citet{resnick2007}, easily extended from $\mathbb{R}_+^p$ to $\mathbb{R}^p$, implies that $ Y = t^{-1}(  X_1)+ t^{-1}(  X_2) \in RV^p(\alpha)$, and has measure $\nu_{ Y}(\cdot)=\nu_{t^{-1}(  X_1)}(\cdot)+\nu_{t^{-1}(  X_2)}(\cdot)$ when normalized by $\{b_n\}$. 
From Lemma~A\ref{lem:tYisRV}, $ X_1 \oplus  X_2= t( Y) \in RV^p_+(\alpha)$, and when normalized by $\{b_n\}$, has measure $\nu_{t( Y)}(\cdot) = \nu_{ Y}\left( \left\{ y\in\bar{\mathbb{R}}^p :  y^{(0)}\in \cdot\right\}\right)$. Further, noting that $\nu_{t^{-1}(  X_1)}$, $\nu_{t^{-1}(  X_2)}$ and thus $\nu_{ Y}$ only have mass on $\bar{\mathbb{X}}^p$, for any $C\subset\bar{\mathbb{X}}^p\setminus\{ 0\}$, $\nu_{ Y}\left( \left\{ y\in\bar{\mathbb{R}}^p :  y^{(0)}\in C\right\}\right)=\nu_{ Y}(C)=\nu_{ X_1}(C) + \nu_{ X_2}(C)$.\smallskip

\noindent {\em Proof of Proposition~\ref{prop: mult}:}  First consider $a t^{-1}( X)$.  For continuity set $C \subset \bar{\mathbb{R}}^p \setminus \{ 0\}$,
\begin{equation}
  \label{eq:lem2half}
  n \Pr[ b_n^{-1}\{a t^{-1}( X)\} \in C ] \rightarrow \nu_{t^{-1}( X)} (a^{-1} C) = \nu_{ X}\{(a^{-1} C) \cap \bar{\mathbb{X}}^p\},
\end{equation}
by Lemma~A\ref{lem:tm1XisRV}.
Hence $a t^{-1}( X) \in RV^p(\alpha)$, and we define $\nu_{a t^{-1}( X)}(\cdot)$ to be its limiting measure when normalized by $\{b_n\}$.
Let $C_+ \subset \bar{\mathbb{X}}^p \setminus \{ 0\}$, and let $C_+^{(0)} = \{  y \in \bar{\mathbb{R}}^p :  y^{(0)} \in C_+\}$.
By Lemma~A\ref{lem:tYisRV} and (\ref{eq:lem2half}),
\begin{equation*}
  n \Pr\{ b_n^{-1}(a \circ  X) \in C_+ \} \rightarrow \nu_{a t^{-1}( X)} \left(C_+^{(0)}\right) =  \nu_{ X}\left\{\left(a^{-1} C_+^{(0)}\right) \cap \bar{\mathbb{X}}^p\right\}.
\end{equation*}
If $a>0$, $(a^{-1} C_+^{(0)}) \cap \bar{\mathbb{X}}^p = a^{-1} (C_+^{(0)} \cap \bar{\mathbb{X}}^p) = a^{-1} C_+$, which implies the first part of the proposition.
If $a \leq 0$, $(a^{-1} C_+^{(0)}) \cap \bar{\mathbb{X}}^p = \emptyset$, as $C_+^{(0)} \cap \{  y \in \bar{\mathbb{R}}^p : y_i \leq 0, i = 1, \ldots p\} = \emptyset$, which implies the second part.\smallskip

We give the following corollary and its proof prior to proving Corollary \ref{coro: mtxMult}.
\begin{corollary}  
  \label{coro: vecMult}
  Let $ a \in \mathbb{R}^p$ where $\max_{i = 1, \ldots, p} a_i > 0$.  Let $Z$ be a regularly varying random variable with index $\alpha$ and assume that $b_n$ is chosen such that $n \Pr(b_n^{-1} Z > x) \rightarrow x^{-\alpha}$, $x>0$. 
  Also assume $n \Pr\{Z \leq \exp(-kn^{1/\alpha})\} \rightarrow 0$ for any $k > 0$.  
  Then $ a \circ Z \in RV_+^p(\alpha)$ and when normalized by $\{b_n\}$ has angular measure 
  $H_{ a \circ Z}(\cdot ) = \|  a^{(0)} \|^\alpha  \delta_{ a^{(0)} / \|  a^{(0)} \|}  (\cdot)$.
\end{corollary}
  
\noindent{\em Proof of Corollary A\ref{coro: vecMult}:} By \citet[Lemma 6.1]{resnick2007}, proof of convergence for sets $[ 0,  x]^c$ for $ x > 0$ which are continuity points of the limit is sufficient for convergence on $M_+(\bar {\mathbb{X}}^p \setminus \{ 0\})$. Note, 
\begin{align*}
  n \Pr\{ b_n^{-1}( a \circ Z) \in [ 0,  x]^c \}
  &= n \Pr \left( \bigcup_{i = 1}^p \left[ t \big\{a_i t^{-1} (Z) \big\} > b_n x_i \right] \right)\\
  &\sim n \Pr \left\{ \bigcup_{i: a_i > 0} \left(Z > {a_i}^{-1}{b_n x_i} \right) \right\} \\
  &= n \Pr \left(   Z > b_n \min_{i: a_i > 0} {a_i}^{-1}{x_i} \right)\\
  &\rightarrow \max_{i = 1, \ldots, p} x_i^{-\alpha}{{a^{(0)}}_i^\alpha}\\
  &= \|  a^{(0)} \|^\alpha \max_{i = 1, \ldots, p} x_i^{-\alpha}{({a^{(0)}}_i/ \|  a^{(0)} \|)^\alpha}.
\end{align*}
Note $\nu_{X} ( [0, x]^c ) = \int_{\mathbb S_{p-1}} \max_{i = 1, \ldots, p} \left( {w_i^\alpha}/{x_i^\alpha} \right) \dd H_{X}(w)$, thus $H_{ a \circ Z}(\cdot ) = \|  a^{(0)} \|^\alpha  \delta_{ a^{(0)} / \|  a^{(0)} \|}  (\cdot)$.\smallskip

{\em Proof of Corollary~\ref{coro: mtxMult}:} As $ a_{j} Z_j$ is independent of $ a_{j'} Z_j'$ for $j \neq j'$, Corollary~\ref{coro: mtxMult} follows from Corollary~A\ref{coro: vecMult} and Proposition~\ref{prop: sum}.\smallskip

The proof of Proposition ~\ref{prop: dense} is in the supplementary materials as it is analogous to the proof found in \cite{fougeres2013}.

\noindent{\em Proof of Proposition~\ref{prop: completelyPositive}:}  By Proposition \ref{prop: dense}, there exists a sequence $\{A_q\}$, $q = 1,2, \ldots$, of nonnegative matrices such that $H_{A_q \circ  Z_q} \stackrel{w}{\rightarrow} H_{ X}$.
For any fixed $q$, let $\Sigma_q = A_q A_q^T$, and note that $\Sigma_q$ is completely positive.
As the set of completely positive matrices is a closed convex cone \citep[Theorem 2.2]{berman2003}, then $\Sigma_{ X} = \lim_{q \rightarrow \infty} \Sigma_q$ is completely positive, and thus there exists a finite $q_*$ and a $p \times q_*$ matrix $A_*$ such that $\Sigma_{ X} = A_{*} A_{*}^T$.  \smallskip

%
%

Lemma A\ref{lem: mtxRegVar} is useful for proving Proposition \ref{prop: pcSigma}.
\begin{lemma}
\label{lem: mtxRegVar}
Let $ X \in RV_+^p(\alpha)$ be such that $n \Pr( b_n^{-1} X \in \cdot ) \stackrel{v}{\rightarrow} \nu_{ X} (\cdot)$. Suppose the matrix $S$ is invertible.
Then $St^{-1}(X) \in RV^p(\alpha)$.
\end{lemma}

\noindent{\em Proof of Lemma~A\ref{lem: mtxRegVar}:}
For any $C\subset\bar{\mathbb{R}}^p\setminus\{ 0\}$ where $S^{-1}C$ is a continuity set of $\nu_{t^{-1}( X)}$,
\begin{equation}
\label{eq:prop5half}
n\Pr[ b_n^{-1} \{S t^{-1}( X)\} \in C ] = n\Pr\{ b_n^{-1} t^{-1}( X) \in S^{-1}C \} \rightarrow \nu_{t^{-1}( X)}(S^{-1}C) = \nu_{t^{-1}( X)}(S^{-1}C \cap \bar{\mathbb{X}}^p),
\end{equation}
where the last equality follows from the fact that $\nu_{t^{-1}( X)}$ only has mass on the positive orthant. 
Hence, $S t^{-1}( X) \in RV^p(\alpha)$.

\noindent{\em Proof of Proposition~\ref{prop: pcSigma}:}
\begin{eqnarray*}
{\sigma_{V}}_{ik} 
&=& \int_{\Theta_{p-1}} v_i v_k \dd H_V(v)\\
&=& \int_{\|y\|_2 \geq 1} \frac{y_i}{\|y\|_2} \frac{y_k}{\|y\|_2} \dd \nu_V(y) \\
&=& \int_{\|U^T x\|_2 \geq 1} \frac{(U^T x)_i}{\|U^T x \|_2} \frac{(U^T x)_k}{\|U^T x \|_2} \dd \nu_X(x) \\
&=& \int_{\|x\|_2 \geq 1} \frac{u_i^T x}{\|x \|_2} \frac{u_k^T x}{\|x \|_2} \dd \nu_X(x); \mbox{ because $U$ is unitary, and where $u_i$ is the $i$th column of $U$} \\
&=& \int_{\Theta^+_{p-1}} (u_i^T w) (w^Tu_k) \dd H_X(w)\\
&=& u_i^T \Sigma_X u_k \\
&=& \left\{ \begin{array}{l}
		\lambda_{i} \mbox{ if } i = k,\\
		0 \mbox{ if } i \neq k.
		\end{array}
	\right.	
\end{eqnarray*}
  
\end{appendix}

\bibliographystyle{apalike}
\bibliography{pcax.bib}

\end{document}